\documentclass[psfig]{aa}
\usepackage{graphics}
\usepackage{psfig}

\begin{document}


\title {Rotational modulation of the photospheric and chromospheric
activity in the young, single K2-dwarf PW And
\thanks{Based on observations made
with the 2.2m telescope of the German-Spanish Astronomical Centre, 
 Calar Alto (Almer\'{\i}a, Spain), 
 operated by the Max-Planck-Institute for Astronomy,
 Heidelberg, jointly with the Spanish National Commission for Astronomy,  
with the Nordic Optical Telescope (NOT),
 operated on the island of La Palma jointly by Denmark, Finland,
 Iceland, Norway and Sweden, in the Spanish Observatorio del 
 Roque de Los Muchachos of the Instituto de Astrof\'{\i}sica de Canarias,
with the Isaac Newton Telescope (INT)
 operated on the island of La Palma by the Isaac Newton Group in
 the Spanish Observatorio del Roque de Los Muchachos of the
 Instituto de Astrof\'{\i}sica de Canarias,
with the Italian Telescopio Nazionale Galileo (TNG) 
 operated on the island of La Palma by the Centro Galileo Galilei of the 
 INAF (Istituto Nazionale di Astrofisica) at the Spanish Observatorio del 
 Roque de Los Muchachos of the Instituto de Astrofisica de Canarias
and with the Hobby-Eberly Telescope (HET)
 operated by McDonald Observatory on behalf 
 of The University of Texas at Austin, the
 Pennsylvania State University, Stanford University, 
 Ludwig-Maximilians-Universit\"at M\"unchen, and
Georg-August-Universit\"at G\"ottingen.
}
\thanks{Tables 1, 3, and 4, and Figures 4, 7, 8, 14, 16, 18, 19, and 21
only available in electronic form
at the CDS via anonymous ftp to {\tt cdsarc.u-strasbg.fr} (130.79.128.5)
or via {\tt http://cdsweb.u-strasbg.fr/cgi-bin/qcat?J/A+A/}}
}

\titlerunning{Rotational modulation of magnetic activity in PW And}

\author{
J.~L\'opez-Santiago\inst{1}
\and D.~Montes\inst{1}
\and M.J.~Fern\'{a}ndez-Figueroa\inst{1}
\and L.W.~Ramsey\inst{2}
}

\offprints{ J.~L\'opez-Santiago}
\mail{jls@astrax.fis.ucm.es}

\institute{
Departamento de Astrof\'{\i}sica,
Facultad de Ciencias F\'{\i}sicas,
 Universidad Complutense de Madrid, E-28040 Madrid, Spain\\
\email{jls@astrax.fis.ucm.es}
\and The Pennsylvania State University, Department of Astronomy and
Astrophysics, 525 Davey Laboratory, University Park, PA 16802, USA
}

\date{Received 20 February 2003 / Accepted 11 August 2003}

\abstract{
High resolution echelle spectra of PW And (HD~1405) have been 
taken during eight observing runs from 1999 to 2002.
The detailed analysis of the spectra allow us to determine its
spectral type (K2V), 
mean heliocentric radial velocity ($V_{\rm hel} = -11.15$ km s$^{\rm -1}$)
rotational velocity ($v\sin{i} = 22.6$ km s$^{\rm -1}$),
and equivalent width of the lithium
line $\lambda 6707.8 \ {\rm \AA}$ ($EW$(Li~{\sc I}) = 273~m\AA).
The kinematic (Galactic Velocity ($U$,$V$,$W$)) 
confirms its membership of the Local Association moving group,
in agreement with the age (30 to 80 Myrs) inferred from the 
color magnitude diagram and the lithium equivalent width.
Photospheric activity (presence of cool spots that disturb the profiles 
of the photospheric lines) has been 
detected as changes in the the bisectors of the 
cross correlation function (CCF) resulting of
cross-correlate the spectra of PW And with the spectrum of
a non active star of similar spectral type.
These variations of the CCF bisectors are related to the 
variations in the measured radial velocities and are modulated 
with a period similar to the photometric period of the star.
At the same time, chromospheric activity has been analyzed, 
using the spectral subtraction technique and simultaneous spectroscopic
observations of the H$\alpha$, H$\beta$, Na~{\sc i} D$_{1}$ 
and D$_{2}$, He~{\sc i} D$_{3}$,
Mg~{\sc i} b triplet, Ca~{\sc ii} H\&K, and Ca~{\sc ii} infrared triplet lines. 
A flare was observed during the last observing run of 2001,
showing an enhancement in the observed chromospheric lines.
A less powerful flare was observed on 2002 August 23.
The variations of the chromospheric activity indicators seem to be related 
to the photospheric activity. 
A correlation between radial velocity, changes in the CCF bisectors 
and equivalent width of different chromospheric lines is observed
with a different behaviour between epochs 1999, 2001 and 2002. 
}
 
\maketitle

\section{Introduction}

In spite of the fact that PW And (HD~1405) is a relatively bright star with a 
visual magnitude ($m_{\rm v}$) of 8.6, it has not been well studied. 
It was included in the Henry Draper Catalogue
as a G5 type star. 
However, other studies show that the spectral type is more
similar to the 
Fehrenbach \& Burnage (1982) classification of PW And as K3 while Christian et al.
(2001) have it as a K0 type star with emission. 
No clear luminosity classification has
been proposed but Hooten \& Hall (1990) suggest it is a main sequence star 
due to its short-period photometric variations ($P \sim 1.75$ days). 
Christian et al. (2001)
identify the EUVE source EUVE J0018+308 as PW And. 
This star was classified as a neighborhood Pleiades-age K2 dwarf 
by Ambruster et al. (1998).
In this work we conclude from its position in the color magnitude diagram 
that PW And could be a pre-main sequence star.
This is consistent with the equivalent width of the lithium
line $\lambda 6707.8 \ {\rm \AA}$ measured by Wichmann et al. (2003)
$EW({\rm Li~{\sc i}}) = 298 \ {\rm m\AA}$ and the abundance of lithium obtained
by Ambruster et al. (1994), $\log N$(Li~{\sc i}) = 3.00 dex.

With a rotational velocity ($v \sin i$) of 23.4 ${\rm km~s^{-1}}$ 
measured by Fekel (1997)
this star is a good candidate for emission in
chromospheric lines like Balmer and Ca~{\sc ii} H \& K lines. 
Bidelman (1985) confirms the
existence of moderate emission in the Ca~{\sc ii} H \& K lines and 
Strassmeier et al. (1988)
include PW And as a candidate in the first catalogue of 
Chromospherically Active Binary Stars.
Nevertheless, studies by Fehrenbach \& Burnage (1982) and Griffin (1992)
fail to confirm a binary nature as the radial velocity ($V_{\rm r}$) 
remains constant with a value of
$-$11.5 and $-$10.5 ${\rm km~s^{-1}}$ respectively.

Space velocity components in addition to its luminosity class, 
rotational velocity, radial velocity as well as its chromospheric
activity all indicate that PW And is a candidate for belonging to 
the Local Association stellar kinematic group 
(for a detailed reference see Montes et al. 2001a, b, 
hereafter M01a and M01b). 

In this paper we are interested primarily in measuring the variations 
caused by magnetic activity in both the photospheric and chromospheric 
lines in the optical spectral region. At the same time we 
carry out a study of the fundamental properties of the star.
This is accomplished using a large set of medium and high resolution echelle 
spectra that were obtained during eight observing runs from 1999 to 2002.  
This interval samples the rotational period during different epochs.
The spectra analyzed here include all the optical 
chromospheric activity indicators
from the Ca~{\sc ii} H \& K to Ca~{\sc ii} IRT lines
as well as the Li~{\sc i} $\lambda$6707.8 line.

In Sect.~2 we give the details of our observations and data reduction.
Stellar parameters determined by us (spectral type, luminosity class, 
rotational and radial velocities and age estimation) are 
given in Sect.~3. In Sect.~4 we present the study of the chromospheric
activity indicators for each observing run, while flare events are 
studied in more detail in Sect.~5. Photospheric variations are
discussed in Sect.~6, 
and Sect.~7 discusses how the photometric variations are related to the 
chromospheric variations. Finally in Sect.~8 we state the conclusions.

\section{Observations and data reduction}
 
The echelle spectra of PW And
analyzed in this paper were obtained during eight observing runs which 
are described in detail in the below paragraphs.

{1)} {2.2m-FOCES 1999/07} \\
This run took place on 
24-29 July 1999 using the 2.2~m telescope at the German Spanish Astronomical
Observatory (CAHA) (Almer\'{\i}a, Spain).
The Fiber Optics Cassegrain Echelle Spectrograph (FOCES)
(Pfeiffer et al. 1998)
was used with a 2048$^{2}$ 15$\mu$ LORAL$\#$11i CCD detector.
The wavelength range covers from
3910 to 9075~\AA$\ $ in 84 orders.
The reciprocal dispersion ranges from 0.03 to 0.07~\AA/pixel
and the spectral resolution,
measured as the full width at half maximum (FWHM)
of the arc comparison lines, ranges from 0.09 to 0.26~\AA.

{2)} {NOT-SOFIN 1999/11} \\
Observations taken on 26-27 November 1999
used the 2.56~m Nordic Optical Telescope (NOT) located
at the Observatorio del Roque de Los Muchachos (La Palma, Spain).
The Soviet Finnish High Resolution Echelle Spectrograph
(SOFIN) was used with an echelle grating (79 grooves/mm),
 Astromed-3200 camera and a 1152$\times$770
pixel EEV P88200 CCD detector. The wavelength range covered is from
3525 to 10425~\AA$\ $ in 44 orders.
The reciprocal dispersion ranges from 0.06 to 0.17~\AA/pixel
and the spectral resolution (FWHM) from 0.14 to 0.32~\AA.

{3)} {INT-MUSICOS 2000/08} \\
Observations made on 5-11 August 2000
with the 2.5~m Isaac Newton Telescope (INT)
at the Observatorio del Roque de Los Muchachos (La Palma, Spain)
used the {ESA-MUSICOS} spectrograph.
This is a fibre-fed cross-dispersed echelle spectrograph,
built as a replica of the
first MUSICOS spectrograph
(Baudrand \& B\"ohm 1992) and developed as part of
MUlti-SIte COntinuous Spectroscopy 
(MUSICOS\footnote{http://www.ucm.es/info/Astrof/MUSICOS.html}) project.
During this observing run,
a 1024$^{2}$ 24$\mu$ TEK5 CCD detector was used,
yielding a wavelength coverage from 4430~\AA$\ $ to 10225~\AA$\ $ in 73 orders.
The reciprocal dispersion ranges from 0.07 to 0.15~\AA$\ $
and the spectral resolution (FWHM) from~0.16 to 0.30~\AA.

{4)} {NOT-SOFIN 2000/11} \\
This run took place on 10-13 November 2000
with the 2.56~m Nordic Optical Telescope (NOT) located
at the Observatorio del Roque de Los Muchachos (La Palma, Spain).
The Soviet Finnish High Resolution Echelle Spectrograph
(SOFIN) was again used in the same configuration as in run (2).

{5)} {2.2m-FOCES 2001/09} \\ 
These observations were made on 21-24 September 
using the 2.2~m telescope at the German Spanish Astronomical
Observatory (CAHA) (Almer\'{\i}a, Spain).
The Fibre Optics Cassegrain Echelle Spectrograph (FOCES)
(Pfeiffer et al. 1998)
was used with a 2048$^{2}$ 24$\mu$ Site$\#$1d CCD detector.
The wavelength range covers from
3510 to 10700~\AA$\ $ in 112 orders.
The reciprocal dispersion ranges from 0.04 to 0.13~\AA/pixel
and the spectral resolution,
determined as the full width at half maximum (FWHM)
of the arc comparison lines, ranges from 0.08 to 0.35~\AA.

{6)} {SARG-TNG 2001/10} \\
Observations were taken on 10-11 October 2001
with the 3.5~m Telescopio Nazionale Galileo (TNG) located
at the Observatorio del Roque de Los Muchachos (La Palma, Spain).
The Spectrografo di Alta Resoluzione Galileo (SARG) 
was used with an echelle grating (31.6 grooves/mm, R4) and the red 
cross-disperser (200 grooves/mm),
and a mosaic of two 2048X4096 CCDs with a 13.5$\mu$ pixel size. 
The wavelength range covers from 4960 to 10110~\AA$\ $ in 62 orders.
The reciprocal dispersion ranges from 0.02 to 0.04~\AA/pixel
and the spectral resolution (FWHM) from 0.08 to 0.17~\AA.

{7)} {HET-HRS 2001/12 - 2002/02} \\
Observations made on 19 December 2001 - 28 February 2002
using the 9.2~m Hobby-Eberly Telescope at McDonald Observatory
in Texas (USA) with the High Resolution Spectrograph (HRS).
The detector is a mosaic of two Marconi 4096 x 2048 CCDs with 15 micron 
pixels (Tull 1998). The wavelength range covers from 5040 to 8775~\AA$\ $
in 52 orders. The reciprocal dispersion ranges from 0.06 to 0.11~\AA/pixel
and the spectral resolution (FWHM) from 0.15 to 0.28~\AA.

{8)} {NOT-SOFIN 2002/08} \\
This took place on 21-29 August 2002
with the 2.56~m Nordic Optical Telescope (NOT) located
at the Observatorio del Roque de Los Muchachos (La Palma, Spain).
The Soviet Finnish High Resolution Echelle Spectrograph
(SOFIN) was used with an echelle grating (79 grooves/mm),
camera 2 and a 2052$\times$2052 
pixel PISKONOV Loral CCD detector. 
The wavelength range covers from 3525 to 10200~\AA$\ $ in 42 orders.
The reciprocal dispersion ranges from 0.015 to 0.045~\AA/pixel
and the spectral resolution (FWHM) from 0.05 to 0.15~\AA.

The velocity resolution derived from the FWHM spectral resolution in the
different observing runs ranges from 4 to 10 km~s$^{-1}$.

\begin{table}
\caption[]{Observing log
\label{tab:obslog}}
\begin{flushleft}
\scriptsize
\begin{tabular}{llccccccccccccc}
\noalign{\smallskip}
\hline
\hline
\noalign{\smallskip}
Name &
\tiny Day/Month  & \tiny UT &  \tiny $S/N$ & \tiny $S/N$ 
\\
     &
 &  &  \tiny (H\&K) & \tiny (H$\alpha$) 
\scriptsize
\\
\noalign{\smallskip}
\hline
\noalign{\smallskip}
{\bf 2.2m-FOCES 1999} \\
\noalign{\smallskip}
\hline
\noalign{\smallskip}
PW And   & 25 Jul & 04:09 & 23 & 123 \\
 "       & 26 "   & 00:57 & 19 & 110 \\
 "       & 27 "   & 02:25 & 13 &  85 \\
 "       & 28 "   & 00:57 & 15 &  76 \\
 "       & 29 "   & 03:01 & 21 & 122 \\
 "       & 30 "   & 01:57 & 20 & 130 \\
GJ 706    & 25 Jul & 00:44 & 37 & 263 \\
 "        & 25 "   & 20:07 & 35 & 176 \\
$\beta$ Oph * & 24 Jul & 23:48 & 31 & 182 \\
 "            & 25 "   & 19:51 & 25 & 222 \\
 "            & 25 "   & 19:55 & 25 & 268 \\
 "            & 27 "   & 20:02 & 22 & 164 \\
 "            & 28 "   & 19:57 & 39 & 332 \\
 "            & 28 "   & 20:01 & 42 & 327 \\
 "            & 29 "   & 19:46 & 32 & 314 \\
\noalign{\smallskip}
\hline
\noalign{\smallskip}
{\bf NOT-SOFIN 1999} \\
\noalign{\smallskip}
\hline
\noalign{\smallskip}
PW And   & 26 Oct & 20:24 & 25 & 107 \\
 "       & 27 "   & 21:29 & 20 & 155 \\
HR 222    *   & 26 Oct & 23:06 & 27 & 119 \\
\noalign{\smallskip}
\hline
\noalign{\smallskip}
{\bf INT-MUSICOS 2000} \\
\noalign{\smallskip}
\hline
\noalign{\smallskip}
PW And   & 11 Aug & 03:36 &  - &  49 \\
 "       & 14 "   & 03:31 &  - &  69 \\
HR 222    *   & 11 Aug & 03:20 &  - & 119 \\
\noalign{\smallskip}
\hline
\noalign{\smallskip}
{\bf NOT-SOFIN 2000} \\
\noalign{\smallskip}
\hline
\noalign{\smallskip}
PW And   & 06 Oct & 01:35 &  7 & 129 \\
 "       & 08 "   & 00:53 & 16 &  89 \\
 "       & 09 "   & 00:03 &  8 & 101 \\
HR 222    *   & 07 Oct & 01:44 & 11 & 234 \\
\noalign{\smallskip}
\hline
\noalign{\smallskip}
{\bf 2.2m-FOCES 2001} \\
\noalign{\smallskip}
\hline
\noalign{\smallskip}
PW And   & 23 Sep & 23:59 & 14 &  91 \\
 "       & 25 "   & 02:03 & 16 &  99 \\
GJ 706   & 23 Sep & 19:52 & 23 & 149 \\ 
HR 166    *   & 23 Sep & 23:50 & 30 & 174 \\
\noalign{\smallskip}
\hline
\noalign{\smallskip}
{\bf TNG-SARG 2001} \\
\noalign{\smallskip}
\hline
\noalign{\smallskip}
PW And   & 12 Oct & 04:05 &  - &  65 \\
HR 222    *   & 12 Oct & 04:30 &  - & 136 \\
\noalign{\smallskip}
\hline
\noalign{\smallskip}
{\bf HET-HRS 2001-2002} \\
\noalign{\smallskip}
\hline
\noalign{\smallskip}
PW And   & 20 Dec & 04:12 &  - & 239 \\
 "       & 21  "  & 03:37 &  - & 338 \\
 "       & 22  "  & 03:44 &  - & 252 \\
 "       & 23  "  & 04:00 &  - & 352 \\
 "       & 26  "  & 03:09 &  - & 184 \\
 "       & 27  "  & 03:07 &  - & 206 \\
 "       & 28  "  & 03:31 &  - & 278 \\
 "       & 29  "  & 02:52 &  - &  67 \\
 "       & 30  "  & 02:59 &  - & 101 \\
HD 92588   *  & 14 Feb  & 09:09 &  - & 246 \\
\noalign{\smallskip}
\hline
\noalign{\smallskip}
{\bf NOT-SOFIN 2002} \\
\noalign{\smallskip}
\hline
\noalign{\smallskip}
PW And   & 22 Aug & 04:16 & 25 & 173 \\
 "       & 23 "   & 04:24 & 21 & 161 \\
 "       & 24 "   & 04:39 & 22 & 145 \\
 "       & 25 "   & 01:55 & 14 & 105 \\
 "       & 26 "   & 05:15 & 14 & 108 \\
 "       & 26 "   & 05:37 & 12 &  77 \\
 "       & 27 "   & 05:06 & 14 & 106 \\
 "       & 28 "   & 04:19 & 19 & 151 \\
 "       & 29 "   & 02:37 & 20 & 138 \\
HR 222     *  & 22 Aug & 06:05 & 24 & 273 \\
 "            & 23 "   & 06:06 & 20 & 302 \\
 "            & 24 "   & 06:13 & 18 & 204 \\
 "            & 25 "   & 02:31 & 32 & 277 \\
 "            & 26 "   & 05:53 & 22 & 218 \\
 "            & 27 "   & 06:04 & 22 & 219 \\
 "            & 28 "   & 05:19 & 15 & 158 \\
 "            & 29 "   & 03:04 & 25 & 182 \\
%
\noalign{\smallskip}
\hline
\end{tabular}

\end{flushleft}
\end{table}

Non-active stars used as reference stars in the spectral subtraction technique
and the radial velocity standards used in the radial velocity determinations
are listed in Tables \ref{tab:obslog} and \ref{tab:par}. 
In Table~\ref{tab:obslog} we give the observing log.
For each observation we list date, UT 
and the signal to noise ratio ($S/N$) obtained in the Ca~{\sc ii}~H~\&~K and
H$\alpha$ line regions. The stars marked with * in the table are standard 
radial velocity stars taken from Beavers et al. (1979).


\begin{table*}
\caption[]{Stellar parameters 
\label{tab:par}}
\begin{flushleft}
\scriptsize
\begin{tabular}{l l c c c c c r r r r }
\noalign{\smallskip}
\hline
\hline
\noalign{\smallskip}
Name & HD& SpT& {$V$--$R$}& {$B$--$V$}& 
$v\sin{i}$ &{$P_{\rm phot}$}& $V_{\rm hel} \pm \sigma_{\rm V_{hel}}$ &
$U\pm \sigma_{\rm U}$ & $V \pm \sigma_{\rm V}$ & $W \pm \sigma_{\rm W}$ \\
     &      &               &       &       &(km s$^{-1}$)&   (days)         
& (km s$^{-1}$) & (km s$^{-1}$) & (km s$^{-1}$) & (km s$^{-1}$) \\
\noalign{\smallskip}
\hline
\noalign{\smallskip}
PW And    & HD 1405    & K2V$^{\rm a}$ & 0.74$^{\rm b}$ & 1.04$^{\rm c}$ 
& 22.6$\pm$0.4$^{\rm a}$  & 1.75$^{\rm d}$ 
& -11.15$\pm$0.05$^{\rm a}$ & -5.42$\pm$0.33$^{\rm a}$
& -28.69$\pm$0.63$^{\rm a}$ &   -17.94$\pm$0.74$^{\rm a}$ \\
\noalign{\smallskip}
\hline
\noalign{\smallskip}
{\bf\tiny Ref. Stars} \\
\noalign{\smallskip}
\hline
\noalign{\smallskip}
HR 4182  *    & HD 92588  & K1IV \\
GJ 706       & HD 166620 & K2V  \\
HR 222   *    & HD 4628   & K2V \\
\hline
\noalign{\smallskip}
\end{tabular}

\end{flushleft}
{\scriptsize
$^{\rm a}$ This work.

$^{\rm b}$ The $V$--$R$ color index are obtained from the
relation with spectral type given by 
Landolt-B\"{o}rnstein (Schmidt-Kaler 1982).

$^{\rm c}$ From Tycho-2 Catalogue (H$\o$g et al. 2000).

$^{\rm d}$ Photometric period calculated by Hooten \& Hall (1990).
}
\end{table*}

The spectra have been extracted using the standard
reduction procedures in the
IRAF\footnote{IRAF is distributed by the National Optical Observatory,
which is operated by the Association of Universities for Research in
Astronomy, Inc., under contract with the National Science Foundation.}
 package (bias subtraction,
flat-field division and optimal extraction of the spectra).
The wavelength calibrations were obtained by taking
spectra of a Th-Ar lamp.
This lamp gives a very rich emission line spectrum over a wide
wavelength range and is thus very useful for high
resolution echelle spectra, including enough calibration lines in each order.
The calibration was performed by fitting all the orders
simultaneously (in both dispersion and cross dispersion directions)
and the residuals obtained in the fit are typically lower than 0.04 \AA.
Finally, the spectra were normalized by using
a cubic spline polynomial fit to the observed continuum.
The points used to fit the continuum in each order are chosen
in regions judged free of spectral lines to reduce as much as
possible the effects of line blending.

\section{Stellar parameters}

\subsection{Spectral type and luminosity class}
To obtain an independent estimate of the spectral type of this star, we have
compared our high resolution echelle spectra with
that of inactive reference stars taken during the same observing run.
The spectral type of the reference stars range from G to M, and 
the luminosity class from {\sc III} to {\sc V}.
%
%
The analysis makes use of the program {\sc starmod}
developed at Penn State University (Barden 1985) and modified more 
recently by us.
With this program a synthetic stellar spectrum is constructed
from the artificially rotationally broadened, and radial-velocity
shifted spectrum of an appropriate reference star.
We obtained the best fit between observed and synthetic spectra
when we use a K2-dwarf spectral type standard star.
The uncertainty in this classification is of one spectral subtype 
as is typical in the MK spectral classification.
In addition, in order to constrain the spectral type and luminosity class,
several temperature-and gravity-sensitive photospheric lines
have been studied in more detail and lead to a very similar result.

\subsection{Radial velocity and space motion}

Heliocentric radial velocities were determined using the 
cross-correlation technique.
The spectra of the star were cross-correlated order by order,
using the routine {\sc fxcor} in IRAF, against spectra of radial velocity
standards of similar spectral types 
(the stars marked with * in Tables \ref{tab:obslog} and \ref{tab:par})
taken from Beavers et al. (1979).
For each order, the velocity is derived 
from the position of the peak of the cross-correlation function (CCF) 
by fitting a Gaussian to the top of the function.
Radial velocity errors are computed by {\sc fxcor} based on the 
fitted peak height and the antisymmetric noise as described by 
Tonry \&  Davis (1979).
The radial velocities calculated for each order are weighted by their 
errors, and a mean value is obtained for each observation 
(see the values given in Table~\ref{tab:vhel} for three representative
 observing runs).
Orders including chromospheric features and prominent telluric lines
were excluded when determining this mean velocity.
Finally, a weighted mean radial velocity is determined using all the 
observations from all the observing runs. 
In Table~\ref{tab:par} we list this average
heliocentric radial velocity (V$_{\rm hel}$)
and its associated error ($\sigma_{\rm V_{hel}}$).

As will be discussed in detail in Section 6 the presence of star-spots 
in the photosphere of this active star can disturb the profile of the 
photospheric lines and induce variations in the peak of the CCF 
which cause variations in the measured radial velocity.
Differences up to 3 ${\rm km~s^{\rm -1}}$ are measured in the 
spectra from the same observing run (see Table~\ref{tab:vhel}).  As
this is larger than the single measurement error these variations 
are attributed to asymmetry variations of the absorption 
lines caused by the stellar activity.
In addition, variations up to 6 ${\rm km~s^{\rm -1}}$
have been found in the radial velocities measured 
at different epochs (from 1999 to 2002).
These variations can be attributed to a) the effects on the line profiles of 
changing levels of stellar activity (such as changes in the spot coverage)
at different epochs, 
b) the use of different radial velocity standard which have different errors,
or c) the changes in the zero-point offset of the different 
observing runs and spectrographs.
However, the final mean radial velocity value we have obtained
($v_{\rm r} = -11.15$ km s$^{-1}$) is very similar to those given by
Fehrenbach \& Burnage (1982) ($v_{\rm r} = -11.5$ km s$^{-1}$)
and Griffin (1992) ($v_{\rm r} = -10.4$ km s$^{-1}$).
Therefore, the radial velocity we report in this paper not only confirms the
previously-measured radial velocities, but allow us to conclusively state
that this star is not a chromospherically active binary, as is suggested by
Strassmeier et al. (1988), since the component
radial velocity amplitude of these kind of binaries
are typically much lager, ranging from 20 to 100 km s$^{-1}$ 
(see Strassmeier et al. 1993).

%

We have used this mean radial velocity together with the spectroscopic parallax 
($M_{\rm V} = 6.4$ from Landolt-B\"orstein (Schmidth-Kaler 1982) for a K2 dwarf 
gives a distance d = 27.54 pc) and proper motions from the Tycho-2 
(H$\o$g et al. 2000) Catalogue, 
to calculate the Galactic space-velocity components ($U$, $V$, $W$) 
in a right-handed coordinated system (positive in the directions of the
Galactic center, Galactic rotation, and the
North Galactic Pole, respectively), as determined 
by M01a. 
The resultant new values are given in Table~\ref{tab:par}.
Its position in the Galactic-velocity diagram 
confirms its membership in the Local Association moving group
(see also M01b and L\'opez-Santiago et al. 2003).
PW And is situated inside the subgroup B4 associated with the
Pleiades cluster identified by Asiain et al. (1999).

\subsection{Rotational velocity}

Fekel (1997) measured the rotational 
velocity ($v \sin i$) of PW And obtaining a value of 
$23.4 \pm 1.0$ km~s$^{\rm -1}$, 
assuming spectral type G5. The mean value from the five Coravel traces is 
$21 \pm 1$ km~s$^{\rm -1}$ and Griffin (1992) determined
$v \sin i = 21.5$ km~s$^{\rm -1}$. 
To determine an accurate rotational velocity of PW And, 
we again make use of the cross-correlation technique
in our high resolution echelle spectra by using the routine
{\sc fxcor} in IRAF.
In each one of the observing runs the observed spectra of PW And were
cross-correlated against the spectrum of the template star
and the width (FWHM) of CCF determined.
The calibration of this width
to yield an estimation of $v\sin{i}$ is done by
cross-correlating artificially broadened spectra of the template star
with the original template star spectrum.
Broadened template spectra are created for $v\sin{i}$ spanning the
expected range of values by convolution with a theoretical rotational
profile (Gray 1992) using the program {\sc starmod}.
The resultant relationship between $v\sin{i}$ and FWHM of
the CCF was then fitted with a fourth-order polynomial.
We have tested this method with stars of known rotational
velocity and have obtained good agreement.
The uncertainties on the $v\sin{i}$ values obtained by this method
have been calculated using the parameter
$R$ defined by Tonry \&  Davis (1979)
as the ratio of the CCF height to the rms
antisymmetric component. 
This parameter is computed by the IRAF task {\sc fxcor}
and provides a measure of the signal to noise ratio of the CCF.
Tonry \&  Davis (1979) show that errors in the FWHM
of the CCF are proportional to $(1 + R)^{-1}$ and
Hartmann et al. (1986) and Rhode et al. (2001) found that the
quantity $\pm v\sin{i}(1 + R)^{-1}$ provides a good estimate
for the 90$\%$ confidence level of a $v\sin{i}$ measurement.
Thus, we have adopted   $\pm v\sin{i}(1 + R)^{-1}$ as a
reasonable estimate of the uncertainties on our $v\sin{i}$ measurements.

We have determined $v\sin{i}$ by this method in all the spectra
available of PW And.
Very similar $v\sin{i}$ values are obtained in spectra on different nights
and at different epochs.
The $v\sin{i}$ derived by this method
is not affected by the asymmetry of the CCF caused by star-spots 
(see Section 6), since
those changes are mainly in the peak of the CCF but 
the $v\sin{i}$ determination 
is based on the FWHM of the CCF.
The resulting error weighted mean 
for all the observing runs is $22.6 \pm 0.4$ km s$^{\rm -1}$, which is the
value given in Table~\ref{tab:par}. Our value is closer to the value
given by Fekel (1997).
The minimum radius ($R\sin{i}$) calculated using our value of $v\sin{i}$ 
and the $P_{\rm phot}$ by Hooten \& Hall (1990) is 
$0.78 \pm 0.11$ R$_{\odot}$, 
which is consistent with a K2 V spectral type. 
Adopting a radius 
$R = 0.80 \pm 0.05$ R$_{\odot}$ 
for a K2 dwarf from the Schmidt-Kaler (1982) tables, 
we obtain the inclination of the rotation axis 
$i = 77 \pm 9^{\small o}$
for PW And.
This is only an estimate, since the radius could be slightly different
due to the possible pre-main sequence nature of this star inferred in the 
next subsection.

\begin{figure}
{\psfig{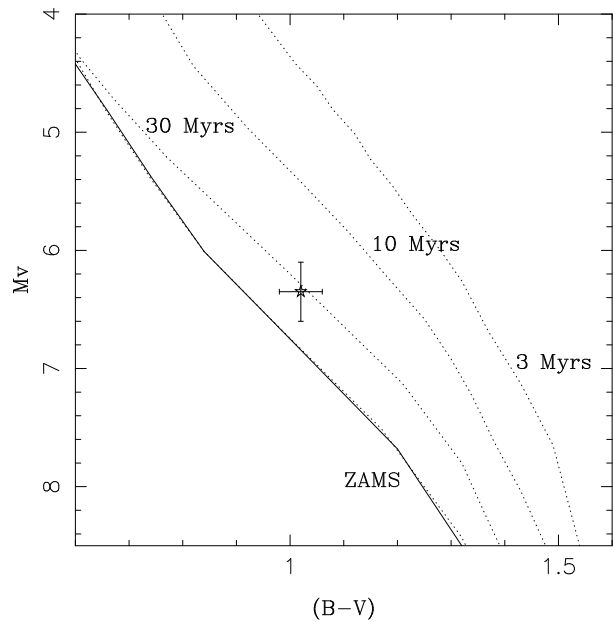}}
\caption[ ]{$M_{\rm v}$-($B - V$) diagram. We have used pre-main sequence
isochrones from Siess et al. (2000).
Continuous line corresponds to ZAMS and dotted lines are 3, 10, 30 and 80
Myrs isochrones (from top to bottom).
\label{fig:isochrone}}
\end{figure}

\subsection{Estimation of age: the color-magnitude diagram}

We estimate the age of PW And from the color-magnitude diagram 
($M_{\rm v}$ vs. ($B$--$V$)).
The value of $M_{\rm v}$ has been adopted from Landolt-B\"ornstein tables
(Schmidt-Kaler 1982) for a dK2 type star. The color index $B$--$V$ 
(listed in Table~\ref{tab:par}) has been obtained from
Tycho-2 Catalogue (H$\o$g et al. 2000) using transformations from $B_{\rm T}$
and $V_{\rm T}$ to Johnson indexes (section 1.3 from Hipparcos Catalogue, 
ESA 1997).
Pre-main sequence isochrones from Siess et al. (2000) have been adopted in 
Fig.~\ref{fig:isochrone} (see L\'opez-Santiago et al. 2003).
The continuous line corresponds to the ZAMS and the dotted lines are 
3, 10, 30 and 80 Myrs isochrones respectively (from top to bottom).
In order to estimate the error bars for $M_{\rm v}$ we have assumed a 
deviation of one spectral type from the dK2 suggested here (K1 to K3). 
Error in $B$--$V$ index has been calculated taking
into account errors in $B_{\rm T}$ and $V_{\rm T}$ given by Tycho-2 and 
their relation with $B$--$V$.
PW And's position in the diagram over the
30 Myrs isochrone (see Fig.~\ref{fig:isochrone}) suggest
that it is a pre-main sequence star. 
Even if we take into account the errors bars (or a large estimation of these) 
the star continue slightly above the ZAMS.

\begin{figure}
{\psfig{figure=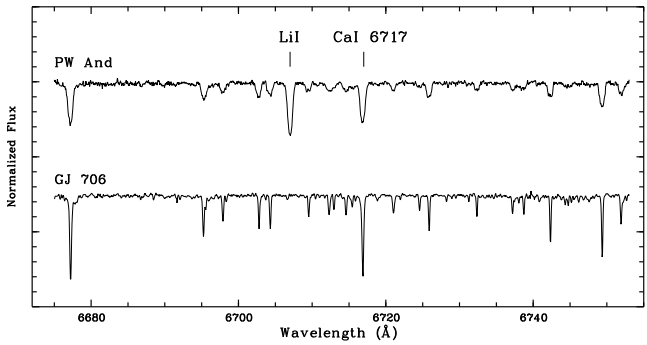,width=8.7cm,bbllx=28pt,bblly=28pt,bburx=550pt,bbury=325pt,clip=}}
\caption[ ]{Spectrum of PW And in the region of Li~{\sc i} $\lambda$6707.8 line.The non-active dK2 star GJ~706 (HD~166620) is plotted as reference.
\label{fig:li}}
\end{figure}

\begin{figure}
{\psfig{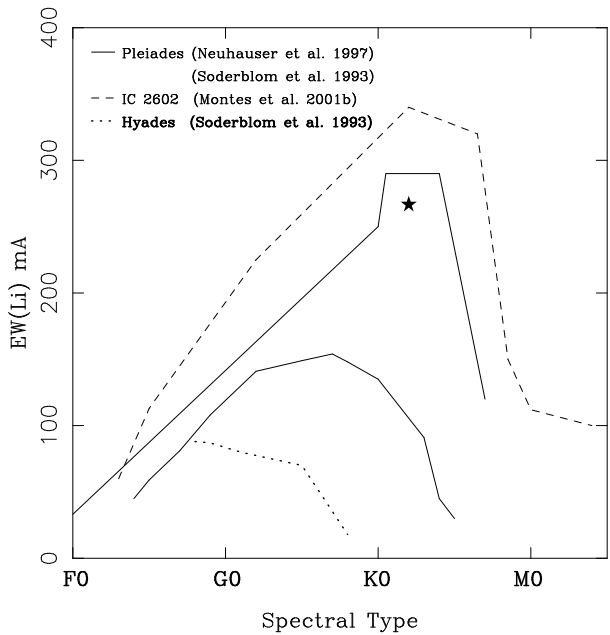}}
\caption[ ]{The figure shows lithium equivalent width ($EW$(Li~{\sc i}))
as a function of spectral type for members of the young open clusters:
IC~2602 (dashed line), Pleiades (solid lines) and Hyades (dotted line).
Only the upper envelope is plotted for IC 2602 and Hyades, while
upper and lower envelopes are presented for Pleiades cluster.
The star corresponds to the position of PW~And in the diagram.
\label{fig:pli}}
\end{figure}

\subsection{Estimation of age: the Li~{\sc i} $\lambda$6707.8 line}

In order to further constrain the age for PW And, the equivalent width of
the resonance doublet of  Li~{\sc i} at $\lambda$6707.8 \AA\ has been measured.
This line is included in our echelle spectra for all the observing runs 
and is an important diagnostic of age in late-type stars since 
the lithium is destroyed easily by thermonuclear reactions in the stellar 
interior.
A representative spectrum in the Li~{\sc i} line region
of the star is plotted in Fig.~\ref{fig:li}.
As can be seen in this figure the Li~{\sc i} absorption line 
is very prominent in PW And in comparison with a reference star 
of the same spectral type.
At this spectral resolution and with the rotational velocity
of the star the Li~{\sc i} line is blended with
the nearby Fe~{\sc i} $\lambda$6707.41~\AA\ line.
We have corrected the total measured equivalent width,
$EW$(Li~{\sc i}+Fe~{\sc i}),
by subtracting the $EW$ of Fe~{\sc i} calculated from the empirical
relationship with ($B$--$V$) given by Soderblom et al. (1990).
The error weighted mean value of the individual values of $EW$(Li~{\sc i}) 
measured on different nights and over different observing runs is  
$266.6 \pm 0.4$ m\AA.  
This value is a bit lower than that measured by Wichmann et al. (2003)
of 273 m\AA\ and corrected for the contribution of the 
Fe~{\sc i} $\lambda$6707.41~\AA\ line. 
Our corrected $EW$(Li~{\sc i}) value 
has been converted into abundances by means of the curves of
growth computed by Pallavicini et al. (1987).
Adopting an effective temperature ($T_{\rm eff}$) of 4900 K  
for PW And, we have obtained 
$\log N$(Li~{\sc i}) = 3.4 dex (on a scale where $\log N({\rm H}) = 12.0$).
This estimation is considerably higher than the obtained by 
Ambruster et al. (1994) of $\log N({\rm Li~{\sc i}}) = 3.0$ dex, 
but these authors do not give any information about the temperature and 
the model they used.

We compare the $EW$(Li~{\sc i}) of PW And
with those of well-known young open clusters of different ages.  
In the $EW$(Li~{\sc i}) versus spectral type diagram (Fig.~\ref{fig:pli})
we have over-plotted the upper envelope of the $EW$(Li~{\sc i}) 
of IC 2602 (10-35~Myr), the Pleiades (78-125~Myr), and the Hyades (600~Myr),
open clusters.
For the Pleiades we adopt the upper envelope determined 
by Neuh\"auser et al. (1997) with data from 
Soderblom et al. (1993) and Garc\'{\i}a L\'{o}pez et al. (1994)
and the lower envelope given by Soderblom et al. (1993).
In the case of IC 2602 we have adopted the upper envelope 
determined in M01b. Finally for the Hyades (600~Myr) 
we have used the upper envelope adopted by Soderblom et al. (1993).

The position of PW And close to the upper envelope of the Pleiades and under 
the upper envelope of IC~2602 (see Fig.~\ref{fig:pli}) suggests an age similar 
to the stars in these clusters. 
Due to the large dispersion in $EW$(Li~{\sc i}) found for stars 
in young clusters,
as in the case of IC 2602 (see M01b), we estimate an age for PW And between
30 to 80 Myrs, corresponding with the ages estimated for the IC 2602 and 
the Pleiades clusters. 
This is consistent with the space motion obtained in section 3.2 and
the age obtained by its position in the color-magnitude diagram 
(Fig.~\ref{fig:isochrone}).

\begin{figure*}
{\psfig{figure=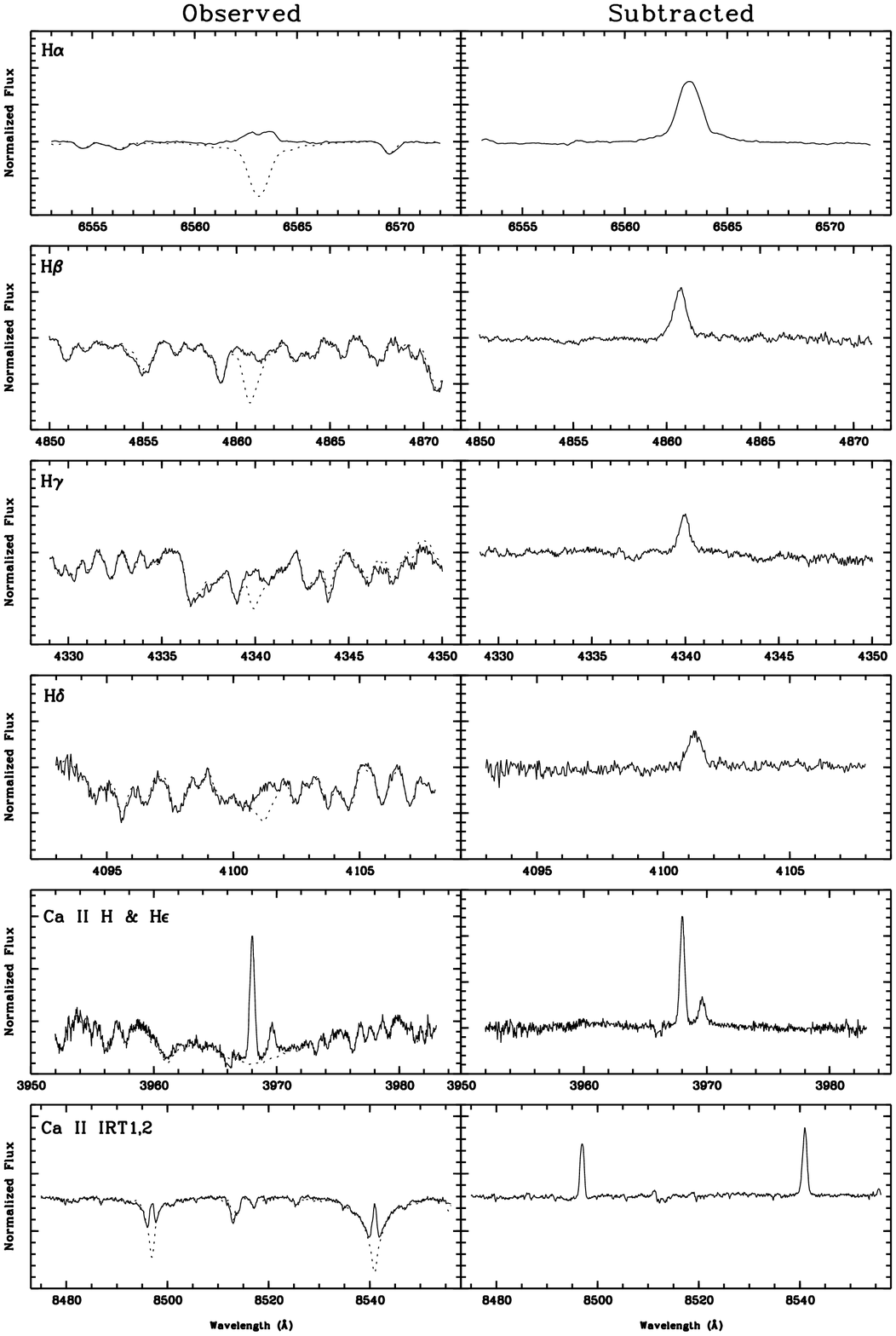,width=18.0cm,clip=}}
\caption[ ]{Representative spectra of PW And in the
quiescent state in the chromospheric activity indicator regions.
\label{fig:activity} }
\end{figure*}

\begin{table*}
\caption[]{EW of the different chromospheric activity indicators
\label{tab:actind}}
\begin{flushleft}
\scriptsize
\begin{tabular}{lcccccccccccc}
\noalign{\smallskip}
\hline
\hline
\noalign{\smallskip}
  &    &   &   & \multicolumn{3}{c}{$EW$(\AA) in the subtracted spectrum} \\
\cline{2-11}
\noalign{\smallskip}
 MJD & \multicolumn{2}{c}{Ca~{\sc ii}} & & & & & &
\multicolumn{3}{c}{Ca~{\sc ii} IRT} \\
\cline{2-3}\cline{9-11}
\noalign{\smallskip}
 & K   & H  & H$\epsilon$ & H$\delta$ & H$\gamma$ & H$\beta$ & H$\alpha$ &
$\lambda$8498 & $\lambda$8542 & $\lambda$8662 \\
\noalign{\smallskip}
\cline{1-11}
\noalign{\smallskip}
\multicolumn{2}{l}{\bf 2.2m-FOCES 1999} \\
\noalign{\smallskip}
\cline{1-11}
\noalign{\smallskip}
51384.1732    & 1.78$\pm$0.09 & 1.17$\pm$0.04 & 0.31$\pm$0.01 & 0.27$\pm$0.01 & 0.29$\pm$0.01 & 0.47$\pm$0.01 & 1.39$\pm$0.04 & 0.49$\pm$0.01 & 0.63$\pm$0.01 & 0.52$\pm$0.01 \\
51385.0401    & 1.36$\pm$0.16 & 0.99$\pm$0.06 & 0.33$\pm$0.02 & 0.28$\pm$0.01 & 0.28$\pm$0.01 & 0.45$\pm$0.01 & 1.26$\pm$0.03 & 0.51$\pm$0.01 & 0.63$\pm$0.01 & 0.56$\pm$0.01 \\
51386.1011    & 1.98$\pm$0.27 & 0.93$\pm$0.05 & 0.32$\pm$0.02 & 0.26$\pm$0.01 & 0.29$\pm$0.01 & 0.51$\pm$0.02 & 1.35$\pm$0.03 & 0.50$\pm$0.01 & 0.66$\pm$0.01 & 0.53$\pm$0.01 \\
51387.0396    & -             & -             & -             & 0.28$\pm$0.02 & 0.27$\pm$0.01 & 0.44$\pm$0.01 & 1.23$\pm$0.03 & 0.47$\pm$0.01 & 0.62$\pm$0.01 & 0.55$\pm$0.01 \\
51388.1258    & 1.97$\pm$0.12 & 0.82$\pm$0.02 & 0.26$\pm$0.01 & 0.24$\pm$0.01 & 0.26$\pm$0.01 & 0.48$\pm$0.01 & 1.20$\pm$0.03 & 0.47$\pm$0.01 & 0.61$\pm$0.01 & 0.52$\pm$0.01 \\
51389.0818    & 2.18$\pm$0.32 & 1.68$\pm$0.03 & 0.33$\pm$0.04 & -             & -             & 0.47$\pm$0.01 & 1.36$\pm$0.04 & 0.49$\pm$0.01 & 0.68$\pm$0.01 & 0.53$\pm$0.01 \\
\noalign{\smallskip}
\cline{1-11}
\noalign{\smallskip}
\multicolumn{2}{l}{\bf NOT-SOFIN 1999} \\
\noalign{\smallskip}
\cline{1-11}
\noalign{\smallskip}
51508.8690    & 2.43$\pm$0.21 & -             & -             & -             & -             & 0.71$\pm$0.02 & 1.58$\pm$0.04 & 0.58$\pm$0.01 & 0.78$\pm$0.06 & -             \\
51509.9022    & 2.36$\pm$0.14 & -             & -             & -             & -             & 0.71$\pm$0.04 & 1.54$\pm$0.04 & 0.53$\pm$0.01 & 0.91$\pm$0.02 & -             \\
\noalign{\smallskip}
\cline{1-11}
\noalign{\smallskip}
\multicolumn{2}{l}{\bf INT-MUSICOS 2000} \\
\noalign{\smallskip}
\cline{1-11}
\noalign{\smallskip}
51767.6572    & -             & -             & -             & -             & 0.17$\pm$0.01 & 0.49$\pm$0.01 & 1.18$\pm$0.02 & 0.49$\pm$0.01 & 0.72$\pm$0.02 & 0.59$\pm$0.02 \\
51770.6541    & -             & -             & -             & -             & -             & -             & 0.84$\pm$0.05 & 0.79$\pm$0.11 & 0.66$\pm$0.06 & 0.89$\pm$0.10 \\
\noalign{\smallskip}
\cline{1-11}
\noalign{\smallskip}
\multicolumn{2}{l}{\bf NOT-SOFIN 2000} \\
\noalign{\smallskip}
\cline{1-11}
\noalign{\smallskip}
51854.5731    & 0.98$\pm$0.19 & -             & -             & 0.20$\pm$0.04 & 0.28$\pm$0.06 & 0.51$\pm$0.02 & 1.39$\pm$0.03 & 0.49$\pm$0.01 & 0.73$\pm$0.02 & -             \\
51855.5441    & 1.20$\pm$0.17 & -             & -             & 0.25$\pm$0.03 & 0.23$\pm$0.03 & 0.54$\pm$0.02 & 1.33$\pm$0.03 & 0.47$\pm$0.01 & 0.76$\pm$0.04 & -             \\
51856.5440    & 1.04$\pm$0.12 & -             & -             & -             & -             & 0.60$\pm$0.02 & 1.20$\pm$0.03 & 0.45$\pm$0.01 & 0.65$\pm$0.03 & -             \\
51857.5090    & 1.19$\pm$0.19 & -           & -               & -             & 0.30$\pm$0.05 & 0.54$\pm$0.02 & 1.48$\pm$0.04 & 0.48$\pm$0.01 & 0.67$\pm$0.02 & -              \\
\noalign{\smallskip}
\cline{1-11}
\noalign{\smallskip}
\multicolumn{2}{l}{\bf 2.2m-FOCES 2001} \\
\noalign{\smallskip}
\cline{1-11}
\noalign{\smallskip}
52176.4998    & 1.79$\pm$0.07 & 1.11$\pm$0.09 & 0.39$\pm$0.03 & 0.29$\pm$0.01 & 0.26$\pm$0.01 & 0.53$\pm$0.01 & 1.25$\pm$0.03 & 0.47$\pm$0.01 & 0.70$\pm$0.01 & 0.55$\pm$0.01 \\
52177.5860    & 1.49$\pm$0.04 & 1.02$\pm$0.06 & 0.33$\pm$0.02 & 0.25$\pm$0.01 & 0.23$\pm$0.01 & 0.47$\pm$0.01 & 1.27$\pm$0.02 & 0.43$\pm$0.01 & 0.63$\pm$0.01 & 0.51$\pm$0.01 \\
\noalign{\smallskip}
\cline{1-11}
\noalign{\smallskip}
\multicolumn{2}{l}{\bf HET-HRS 2001-2002} \\
\noalign{\smallskip}
\cline{1-11}
\noalign{\smallskip}
52263.6771    & -             & -             & -             & -             & -             & -             & 1.26$\pm$0.02 & 0.43$\pm$0.01 & 0.71$\pm$0.02 & 0.64$\pm$0.02 \\
52264.6541    & -             & -             & -             & -             & -             & -             & 1.16$\pm$0.02 & 0.50$\pm$0.02 & 0.64$\pm$0.03 & 0.55$\pm$0.02 \\
52265.6573    & -             & -             & -             & -             & -             & -             & 1.31$\pm$0.02 & 0.48$\pm$0.01 & 0.74$\pm$0.02 & 0.56$\pm$0.01 \\
52266.6686$^{\rm a}$    & -             & -             & -             & -             & -             & -             & 2.75$\pm$0.04 & 0.67$\pm$0.02 & 1.03$\pm$0.02 & 0.76$\pm$0.02 \\
52269.6331    & -             & -             & -             & -             & -             & -             & 1.42$\pm$0.02 & 0.44$\pm$0.01 & 0.74$\pm$0.02 & 0.68$\pm$0.02 \\
52270.6320    & -             & -             & -             & -             & -             & -             & 1.43$\pm$0.02 & 0.52$\pm$0.01 & 0.72$\pm$0.02 & 0.60$\pm$0.03 \\
52271.6485    & -             & -             & -             & -             & -             & -             & 1.32$\pm$0.03 & 0.43$\pm$0.01 & 0.72$\pm$0.02 & 0.57$\pm$0.01 \\
52272.6263    & -             & -             & -             & -             & -             & -             & 1.15$\pm$0.03 & 0.44$\pm$0.02 & -             & 0.42$\pm$0.02 \\
52273.6263    & -             & -             & -             & -             & -             & -             & 1.28$\pm$0.04 & 0.42$\pm$0.01 & 0.73$\pm$0.02 & 0.54$\pm$0.01 \\
\noalign{\smallskip}
\cline{1-11}
\noalign{\smallskip}
\multicolumn{2}{l}{\bf NOT-SOFIN 2002} \\
\noalign{\smallskip}
\cline{1-11}
\noalign{\smallskip}
52508.6869    & 2.79$\pm$0.14 & -             & -             & -             & -             & 0.49$\pm$0.01 & 1.31$\pm$0.02 & -             & -             & -             \\
52509.6923$^{\rm a}$    & 3.01$\pm$0.18 & -             & -             & -             & -             & 0.68$\pm$0.02 & 1.71$\pm$0.03 & -             & -             & -             \\
52510.7007    & 2.48$\pm$0.15 & -             & -             & -             & -             & 0.49$\pm$0.01 & 1.29$\pm$0.01 & -             & -             & -             \\
52511.5869    & 2.08$\pm$0.20 & -             & -             & -             & -             & 0.54$\pm$0.01 & 1.25$\pm$0.02 & -             & -             & -             \\
52512.7255    & 2.21$\pm$0.20 & -             & -             & -             & -             & 0.46$\pm$0.01 & 1.17$\pm$0.01 & -             & -             & -             \\
52512.7377$^{\rm b}$    & 1.57$\pm$0.18 & -            & -             & -             & -             & 0.44$\pm$0.01 & 1.21$\pm$0.02 & -             & -             & -             \\
52513.7195    & 2.06$\pm$0.19 & -             & -             & -             & -             & 0.44$\pm$0.01 & 1.29$\pm$0.01 & -             & -             & -             \\
52514.6870    & 2.40$\pm$0.14 & -             & -             & -             & -             & 0.66$\pm$0.01 & 1.42$\pm$0.02 & -             & -             & -             \\
52515.6157    & 2.24$\pm$0.18 & -             & -             & -             & -             & 0.62$\pm$0.02 & 1.35$\pm$0.02 & -             & -             & -             \\
\noalign{\smallskip}
\cline{1-11}
\end{tabular}
\end{flushleft}
{\scriptsize
$^{\rm a}$ Flare detected.

$^{\rm b}$ Low S/N obtained.
}

\end{table*}

\begin{table*}
\caption[]{Absolute surface flux
of the different chromospheric activity indicators
\label{tab:actflux}}
\begin{flushleft}
\scriptsize
\begin{tabular}{lcccccccccccccc}
\noalign{\smallskip}
\hline
\hline
\noalign{\smallskip}
     &      &  &   & \multicolumn{3}{c}{logF$_{\rm S}$ (erg cm$^{-2}$ s$^{-1}$)} & & & & & & 
\multicolumn{3}{c}{Excess Emission}\\
\cline{2-11}\cline{13-15}
\noalign{\smallskip}
 MJD & \multicolumn{2}{c}{Ca~{\sc ii}} & & & & & &
\multicolumn{3}{c}{Ca~{\sc ii} IRT}\\
\cline{2-3}\cline{9-11}
\noalign{\smallskip}
 & K   & H  & H$\epsilon$ & H$\delta$ & H$\gamma$ & H$\beta$ & H$\alpha$ &
$\lambda$8498 & $\lambda$8542 & $\lambda$8662 & & 
$\frac{EW({\rm H\alpha})}{EW({\rm H\beta})}$ & 
$\frac{E_{\rm H\alpha}}{E_{\rm H\beta}}$ & $\frac{E_{8542}}{E_{8498}}$
\scriptsize
\\
\noalign{\smallskip}
\cline{1-15}
\noalign{\smallskip}
\multicolumn{2}{l}{\bf 2.2m-FOCES 1999} \\
\noalign{\smallskip}
\cline{1-15}
\noalign{\smallskip}
51384.1732 & 6.397 & 6.214 & 5.638 & 5.608 & 5.682 & 5.986 & 6.557 & 6.061 & 6.170 & 6.087 & & 2.96 & 3.72 & 1.29\\
51385.0401 & 6.280 & 6.142 & 5.665 & 5.624 & 5.667 & 5.967 & 6.514 & 6.078 & 6.170 & 6.119 & & 2.80 & 3.53 & 1.24\\
51386.1011 & 6.443 & 6.115 & 5.651 & 5.592 & 5.682 & 6.021 & 6.544 & 6.070 & 6.190 & 6.095 & & 2.65 & 3.33 & 1.32\\
51387.0396 &   -   &   -   &   -   & 5.624 & 5.651 & 5.957 & 6.504 & 6.043 & 6.163 & 6.111 & & 2.80 & 3.52 & 1.32\\
51388.1258 & 6.441 & 6.060 & 5.561 & 5.557 & 5.635 & 5.995 & 6.493 & 6.043 & 6.156 & 6.087 & & 2.50 & 3.15 & 1.30\\
51389.0818 & 6.485 & 6.372 & 5.665 &  -    &  -    & 5.986 & 6.547 & 6.061 & 6.203 & 6.095 & & 2.89 & 3.64 & 1.39\\
\noalign{\smallskip}
\cline{1-15}
\noalign{\smallskip}
\multicolumn{2}{l}{\bf NOT-SOFIN 1999} \\
\noalign{\smallskip}
\cline{1-15}
\noalign{\smallskip}
51508.8690 & 6.532 &  -    &  -    &  -    &  -    & 6.165 & 6.612 & 6.134 & 6.263 &  -    & & 2.23 & 2.80 & 1.34\\
51509.9022 & 6.519 &  -    &  -    &  -    &  -    & 6.165 & 6.601 & 6.095 & 6.330 &  -    & & 2.17 & 2.73 & 1.72\\
\noalign{\smallskip}
\cline{1-15}
\noalign{\smallskip}
\multicolumn{2}{l}{\bf INT-MUSICOS 2000} \\
\noalign{\smallskip}
\cline{1-15}
\noalign{\smallskip}
51767.6572 &  -    &  -    &  -    &  -    & 5.450 & 6.004 & 6.486 & 6.061 & 6.228 & 6.141 & & 2.41 & 3.03 & 1.47\\
51770.6541 &  -    &  -    &  -    &  -    &  -    &  -    & 6.338 & 6.268 & 6.190 & 6.320 & &  -   &  -   & 0.84\\
\noalign{\smallskip}
\cline{1-15}
\noalign{\smallskip}
\multicolumn{2}{l}{\bf NOT-SOFIN 2000} \\
\noalign{\smallskip}
\cline{1-15}
\noalign{\smallskip}
51854.5731 & 6.137 &  -    &  -    & 5.478 & 5.667 & 6.021 & 6.557 & 6.061 & 6.234 &  -    & & 2.73 & 3.43 & 1.49\\
51855.5441 & 6.225 &  -    &  -    & 5.574 & 5.581 & 6.046 & 6.538 & 6.043 & 6.251 &  -    & & 2.46 & 3.10 & 1.62\\
51856.5440 & 6.163 &  -    &  -    &  -    &  -    & 6.092 & 6.493 & 6.024 & 6.184 &  -    & & 2.00 & 2.52 & 1.44\\
51857.5090 & 6.222 &  -    &  -    &  -    & 5.697 & 6.046 & 6.584 & 6.052 & 6.197 &  -    & & 2.74 & 3.45 & 1.40\\
\noalign{\smallskip}
\cline{1-15}
\noalign{\smallskip}
\multicolumn{2}{l}{\bf 2.2m-FOCES 2001} \\
\noalign{\smallskip}
\cline{1-15}
\noalign{\smallskip}
52176.4998 & 6.399 & 6.192 & 5.737 & 5.639 & 5.635 & 6.038 & 6.511 & 6.043 & 6.216 & 6.111 & & 2.36 & 2.97 & 1.49\\
52177.5860 & 6.319 & 6.155 & 5.665 & 5.574 & 5.581 & 5.986 & 6.518 & 6.004 & 6.170 & 6.078 & & 2.70 & 3.40 & 1.47\\
\noalign{\smallskip}
\cline{1-15}
\noalign{\smallskip}
\multicolumn{2}{l}{\bf HET-HRS 2001-2002} \\
\noalign{\smallskip}
\cline{1-15}
\noalign{\smallskip}
52263.6771 &  -    &  -    &  -    &  -    &  -    &  -    & 6.514 & 6.004 & 6.222 & 6.177 &  &  -  & -   & 1.65\\
52264.6541 &  -    &  -    &  -    &  -    &  -    &  -    & 6.478 & 6.070 & 6.177 & 6.111 &  &  -  & -   & 1.28\\
52265.6573 &  -    &  -    &  -    &  -    &  -    &  -    & 6.531 & 6.052 & 6.240 & 6.119 &  &  -  & -   & 1.54\\
52266.6686$^{\rm a}$ &  -    &  -    &  -    &  -    &  -    &  -    & 6.853 & 6.197 & 6.383 & 6.251 &  &  -  & -   & 1.54\\
52269.6331 &  -    &  -    &  -    &  -    &  -    &  -    & 6.566 & 6.014 & 6.240 & 6.203 &  &  -  & -   & 1.68\\
52270.6320 &  -    &  -    &  -    &  -    &  -    &  -    & 6.569 & 6.087 & 6.228 & 6.149 &  &  -  & -   & 1.38\\
52271.6485 &  -    &  -    &  -    &  -    &  -    &  -    & 6.534 & 6.004 & 6.228 & 6.126 &  &  -  & -   & 1.67\\
52272.6263 &  -    &  -    &  -    &  -    &  -    &  -    & 6.474 & 6.014 &  -    & 5.994 &  &  -  & -   &  -  \\
52273.6263 &  -    &  -    &  -    &  -    &  -    &  -    & 6.521 & 5.994 & 6.234 & 6.103 &  &  -  & -   & 1.74\\
\noalign{\smallskip}
\cline{1-15}
\noalign{\smallskip}
\multicolumn{2}{l}{\bf NOT-SOFIN 2002} \\
\noalign{\smallskip}
\cline{1-15}
\noalign{\smallskip}
52508.6869 & 6.592 &  -    &  -    &  -    &  -    & 6.004 & 6.531 &  -    &  -    &  -    & & 2.67 & 3.37 &  -  \\
52509.6923$^{\rm a}$ & 6.625 &  -    &  -    &  -    &  -    & 6.146 & 6.647 &  -    &  -    &  -    & & 2.51 & 3.17 &  -  \\
52510.7007 & 6.541 &  -    &  -    &  -    &  -    & 6.004 & 6.524 &  -    &  -    &  -    & & 2.63 & 3.32 &  -  \\
52511.5869 & 6.464 &  -    &  -    &  -    &  -    & 6.046 & 6.511 &  -    &  -    &  -    & & 2.31 & 2.92 &  -  \\
52512.7255 & 6.491 &  -    &  -    &  -    &  -    & 5.976 & 6.482 &  -    &  -    &  -    & & 2.54 & 3.20 &  -  \\
52512.7377$^{\rm b}$ & 6.342 &  -    &  -    &  -    &  -    & 5.957 & 6.497 &  -    &  -    &  -    & & 2.75 & 3.46 &  -  \\
52513.7195 & 6.460 &  -    &  -    &  -    &  -    & 5.957 & 6.524 &  -    &  -    &  -    & & 2.93 & 3.69 &  -  \\
52514.6870 & 6.526 &  -    &  -    &  -    &  -    & 6.133 & 6.566 &  -    &  -    &  -    & & 2.15 & 2.71 &  -  \\
52515.6157 & 6.496 &  -    &  -    &  -    &  -    & 6.106 & 6.544 &  -    &  -    &  -    & & 2.18 & 2.74 &  -  \\
\noalign{\smallskip}
\cline{1-15}
\end{tabular}
\end{flushleft}
{\scriptsize
$^{\rm a}$ Flare detected.

$^{\rm b}$ Low S/N obtained.
}

\end{table*}

\section{Chromospheric activity indicators}

The activity of PW And was previously studied by
Bidelman (1985) who 
reported moderate Ca~{\sc ii} H \& K emission
and the H$\alpha$ line in emission. 
Chromospheric and transition region UV emission fluxes have been reported
by Ambruster et al. (1998) and Wood et al. (2000).
In addition, it has been detected by the ROSAT-satellite as the
2RE J001820+305 source
(Pye et al. 1995; Kreysing et al. 1995; Thomas et al. 1998;
Rutledge et al. 2000),
and by the EUVE-satellite as the EUVE J0018+309 source
(Malina et al. 1994; Christian et al. 2001).

The echelle spectra analyzed in this paper allow us to study
the behaviour of different optical
chromospheric activity indicators
from the Ca~{\sc ii} H \& K to the Ca~{\sc ii} IRT lines,
which are formed at different atmospheric heights. 
Using the spectral subtraction technique described 
by Montes et al. (1995; 1997; 1998; 2000)
it is possible to study in detail the chromosphere, discriminating between
different structures: plages, prominences, flares and microflares.
The synthesized spectra were constructed using the program {\sc starmod}
developed at Penn State University (Barden 1985) and modified by us. 
The inactive stars used as reference stars in the spectral subtraction
were observed during the same observing run than the active stars 
(listed in Table~\ref{tab:obslog}).
Representative spectra in the Ca~{\sc ii} H \& K, H$\delta$, H$\gamma$, 
H$\beta$, H$\alpha$ and Ca~{\sc ii} IRT ($\lambda$8498, $\lambda$8542) 
line regions
of the quiescent state of PW And are presented in Fig.~\ref{fig:activity}.
For each region we have plotted the observed spectrum (solid-line) and the
synthesized spectrum (dashed-line) in the left panel
and the subtracted spectrum in the right panel.
H$\alpha$ emission above the continuum is detected showing an 
autoabsorption feature.
Filled-in absorption in other Balmer lines is also detected, 
and strong emission is observed in Ca~{\sc ii}~H~\&~K, H$_{\epsilon}$ and 
Ca~{\sc ii}~IRT lines.

The equivalent width ($EW$) (measured in the
subtracted spectra) for the Ca~{\sc ii} H \& K, H$\epsilon$,
H$\delta$, H$\gamma$, H$\beta$, H$\alpha$, and  Ca~{\sc ii} IRT
($\lambda$8498, $\lambda$8542, $\lambda$8662) lines for
each observation is given in Table~\ref{tab:actind}.
An estimate of the error for the $EW$ can be obtained from the relation 
given by Cayrel (1988):

\begin{equation}
\Delta EW_{\rm \lambda} = \frac{1.6\sqrt{\omega \delta x}}{S/N}
\end{equation}

where $\omega$ is the FWHM of the line, $\delta x$ the pixel size in \AA, 
and $S$/$N$ the signal-to-noise ratio per pixel in the continuum. 
With a typical $\omega$ of 1.35 \AA\ for 
the H$\alpha$ line, $\delta x = 0.09$ \AA\ and 
$S/N \sim 100$ for our observations, the error estimate 
is in the order of 6 m\AA. This equation does not take into account errors in 
the normalization, as well as possible blends, and must be used with the 
observed spectrum instead of the subtracted one. 
We have estimated the errors in the measured $EW$ 
(see Table~\ref{tab:actind}) taking into account
the rms error obtained from the fit between observed and
synthesized spectra in the regions outside the chromospheric features
and the standard deviations resulting from the
$EW$ measurements. Deviations in the normalization, differences between the
active and the reference stars, and internal precisions of {\sc starmod}
(0.5 - 2 km s$^{-1}$ in velocity shifts, and
$\pm$ 5 km s$^{-1}$ in  $v\sin{i}$) are taken into account 
by this method.
Errors in the chromospheric features in the blue spectral region are 
larger due to the lower S/N of the spectra in this region. 
As an indication of the accuracy of the data, we give in
Table~\ref{tab:obslog} the S/N in 
the Ca~{\sc ii} H \& K, and H$\alpha$ line regions.

\begin{figure}[t]
{\psfig{figure=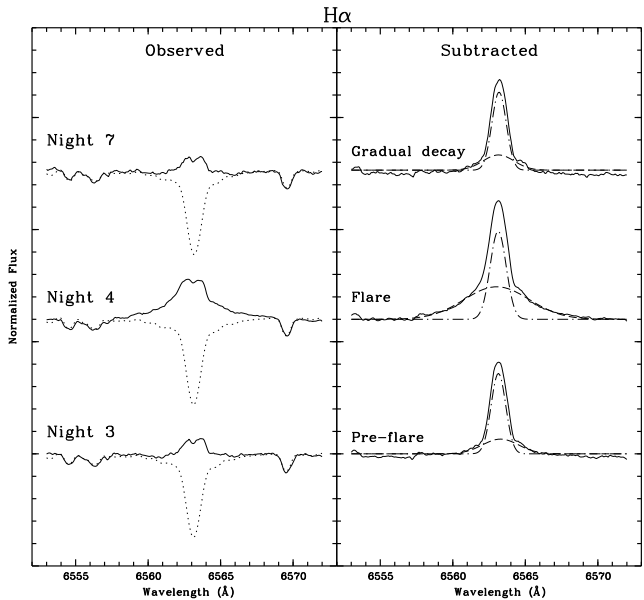,width=9.0cm,bbllx=28pt,bblly=28pt,bburx=570pt,bbury=525pt,clip=}}
\caption[ ]{Comparison between pre-flare (night~3), flare maximum (night~4)
and gradual decay (night~7) in H$\alpha$ line in the observing run
HET-HRS 2001/12.
Dashed lines on the left panel correspond to the reference star.
Dashed and dotted-dashed lines on the right panel
are for broad (B) and narrow (N) components respectively.
\label{fig:flare_ha_het}}
\end{figure}

\begin{figure}[t]
{\psfig{figure=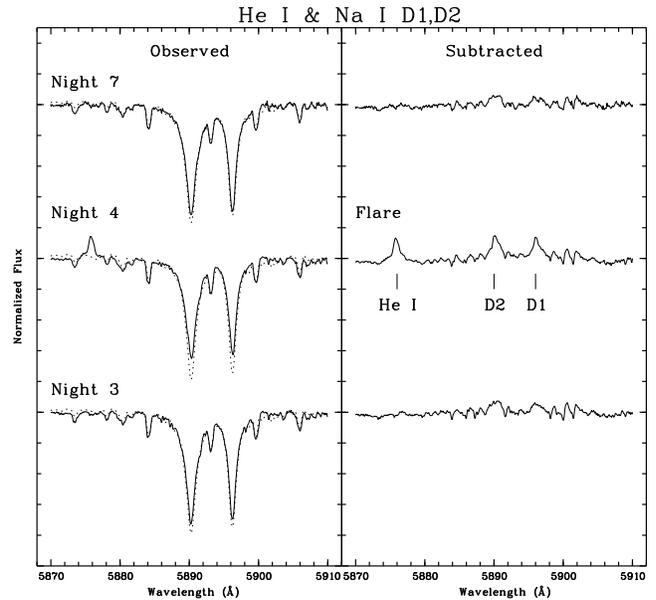,width=9.0cm,bbllx=28pt,bblly=28pt,bburx=570pt,bbury=525pt,clip=}}
\caption[ ]{Comparison between pre-flare (night~3), flare maximum (night~4)
and gradual decay (night~7) in He~{\sc i} D$_{\rm 3}$ and
Na~{\sc i} (D$_{\rm 1}$,D$_{\rm 2}$) lines which are present as emission in the
subtracted spectrum at flare maximum in the observing run
HET-HRS 2001/12.
Dashed lines on the left panel correspond to the reference star.
\label{fig:flare_na_het}}
\end{figure}

The EW have been converted to absolute chromospheric flux at
the stellar surface by using the calibration
of Hall (1996) as a function of ($B$--$V$).
In Table~\ref{tab:actflux}
we give the absolute flux at the stellar surface
(log$F_{\rm S}$) for the lines listed in
Table~\ref{tab:actind}.

In Table~\ref{tab:actflux} we also give the ratio of excess emission $EW$,
$\frac{EW({\rm H\alpha})}{EW({\rm H\beta})}$, 
for all the observations of PW And, 
and the ratio of the excess emission $\frac{E_{8542}}{E_{8498}}$, and 
$\frac{E_{\rm H\alpha}}{E_{\rm H\beta}}$ with the correction:

\begin{equation}
\frac{E_{\rm H\alpha}}{E_{\rm H\beta}} = \frac{EW({\rm H\alpha})}{EW({\rm H\beta})} * 0.2444 * 2.512^{B-R}
\end{equation}

given by Hall \& Ramsey (1992) that takes into account the absolute 
flux density in these lines
and the color difference in the components.

We have used this ratio as a diagnostic for discriminating between 
the presence of plages and prominences on the stellar surface, 
following the results of Hall \& Ramsey (1992)
who found that low $E_{\rm H\alpha}$/$E_{\rm H\beta}$ ($\sim$ 1-2) 
can be achieved both in plages and prominences viewed against the disk, 
but that high ratios ($\sim$ 3-15) can only
be achieved in extended regions viewed off the limb. The high ratio 
($E_{\rm H\alpha}$/$E_{\rm H\beta}$$>$3, see Table~\ref{tab:actind}) 
that we have found in PW And indicates that the emission would arise 
from extended regions (prominences).
$E_{8542}$/$E_{8498}$ ratios are in the range~$\sim$~1-2 which is indicative 
of optically thick emission in plage-like regions, in contrast with the 
prominence-like material inferred by the 
$E_{\rm H\alpha}$/$E_{\rm H\beta}$ ratios.

\section{Flare state}

Optical flares are commonly observed in dMe stars, however, 
in more luminous stars flares are usually only detected through UV or X-ray 
observations (e.g., Landini et al.  1986; H\"unsch \& Reimers 1995; 
Ayres et al. 1994). Optical flares are rare for K-type 
stars and only a few of them have been reported. 
Montes et al. (1999) detected a strong 
optical flare on LQ Hya, a K2 dwarf star very similar to PW And, 
with a duration of at least 5 hours.

Abbott et al. (1995) reported a photometric flare on PW And in the U band 
($\Delta U \sim 0.2$).
Two optical flares are detected on PW And in our data set. 
The first one was observed during the fourth night of the HET/HRS 12/2001 
observing run (2001 December 23), 
and exhibited an enhancement in both H$\alpha$ and the 
Ca~{\sc ii} IRT lines. The second flare was observed during the
NOT-SOFIN 2002/08 observing run (2002 August 23) and while it appears to be a 
less powerful event it is possible that the flare maximum may have occurred 
between observations.

\begin{figure}[t]
{\psfig{figure=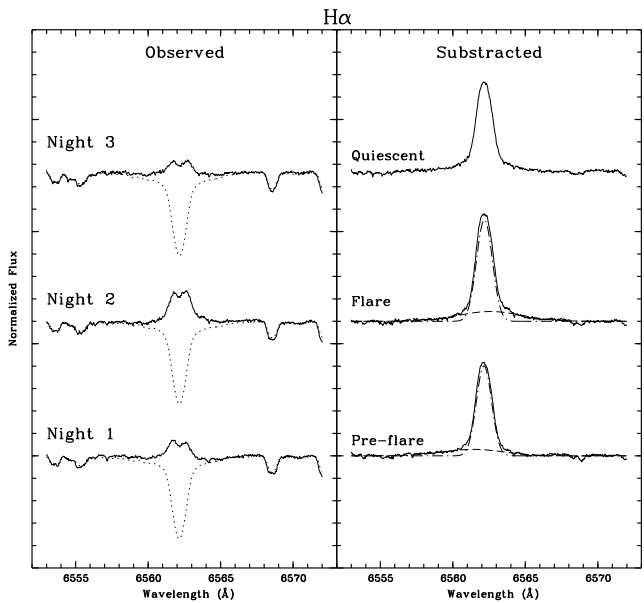,width=9.0cm,bbllx=28pt,bblly=28pt,bburx=570pt,bbury=525pt,clip=}}
\caption[ ]{Comparison between pre-flare (night~1), flare maximum (night~2)
and quiescent (night~3) in H$\alpha$ line in the observing run
NOT-SOFIN 2002/08.
Dashed lines on the left panel correspond to the reference star.
Dashed and dotted-dashed lines on the right panel
are for broad (B) and narrow (N) components respectively.
\label{fig:flare_ha_not}}
\end{figure}

\begin{figure}[t]
{\psfig{figure=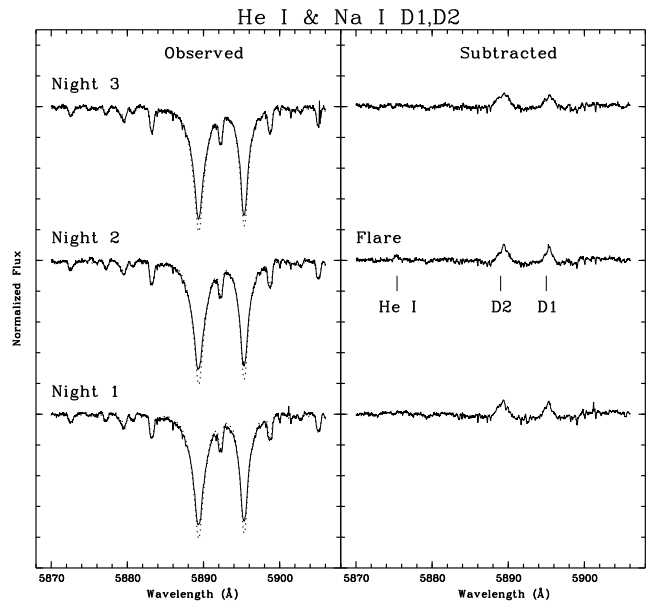,width=9.0cm,bbllx=28pt,bblly=28pt,bburx=570pt,bbury=525pt,clip=}}
\caption[ ]{Comparison between pre-flare (night~1), flare maximum (night~2)
and quiescent (night~3) in He~{\sc i} D$_{\rm 3}$ and
Na~{\sc i} (D$_{\rm 1}$,D$_{\rm 2}$) lines which are present as emission in the
subtracted spectrum at flare maximum in the observing run
NOT-SOFIN 2002/08.
Dashed lines on the left panel correspond to the reference star.
\label{fig:flare_na_not}}
\end{figure}

\subsection{Description of the flare on 2001 December 23}

The H~{\sc i} Balmer lines are in emission in the observed 
and subtracted spectra at 
flare maximum and quiescent state as can be readily seen 
in Fig.~\ref{fig:flare_ha_het} . 
On the other hand, Na~{\sc i} D$_{\rm 1}$ and D$_{\rm 2}$ lines
are clearly revealed only in the subtracted spectra. 
The usual chromospheric flare diagnostic 
lines He~{\sc i}~D$_{\rm 3}$ and the Mg~{\sc i}~b triplet are only 
observed at flare maximum.
Observed and subtracted spectra in the region of 
He~{\sc i} D$_{\rm 3}$ and Na~{\sc i} 
doublet are plotted in Fig.~\ref{fig:flare_na_het} 
for the quiescent state (night~3, 2001 Dec 22), 
flare maximum (night~4, 2001 Dec 23)
and gradual decay (night~7, 2001 Dec 26).

The H$\alpha$ emission equivalent width ($EW$) in the subtracted
spectra, increases a factor of $\sim$~2.1 from the third to the fourth 
nights of the observing run 
(see Fig.~\ref{fig:flare_ha_het} and Table~\ref{tab:actind}), 
while the Ca~{\sc ii} IRT lines increases a factor of $\sim$ 1.4
(see Table~\ref{tab:actind}).
In Table~\ref{tab:hel_flare} we list the measured 
$EW$ of He~{\sc i} D$_{\rm 3}$, Na~{\sc i} doublet and Mg~{\sc i} b triplet 
in the subtracted spectra for the flare spectra and the presumed quiescent 
spectra.
Emission in the Na~{\sc i} doublet is observed in the subtracted spectra 
throughout the flare event, being stronger at flare maximum, 
(see Fig.~\ref{fig:flare_na_het}) while emission 
in He~{\sc i} D$_{\rm 3}$ and 
Mg~{\sc i} b triplet is present only at the observed flare maximum.
The detection of prominent He~{\sc i} D$_{\rm 3}$ emission 
indicates that we are observing an energetic flare in PW And, 
as this feature is 
typically only observed in flare events in very active stars such 
as RS CVn systems and UV Ceti flare stars 
(see Montes et al. 1999 and references therein).
The Fe~{\sc ii} $\lambda$5169~\AA \ line near the Mg~{\sc ii}~b triplet 
is also observed in emission at the flare maximum 
(see Table~\ref{tab:hel_flare}). 

One notable feature is the existence of a broad emission component in both 
the subtracted and observed profiles in H$\alpha$ at the flare maximum.
A two Gaussian fit is necessary to obtain a good 
match to the total emission: a narrow component (N) having a FWHM of 
$\sim$ 59 km~s$^{\rm -1}$ and a broad component (B) with FWHM 
$\sim$ 242 km~s$^{\rm -1}$. 
The profile is also asymmetric with the B component blue-shifted. This
could be interpreted as high velocity mass ejection (Houdebine et al. 1990) 
or high velocity chromospheric evaporation (Gunn et al. 1994).
A broad emission component is also detected in 
the H$\alpha$ subtracted profile during the gradual decay, yielding a FWHM 
of 53 km s$^{\rm -1}$ and 115 km s$^{\rm -1}$ respectively for the 
N and B components.
Again, the profile is asymmetric with the B component showing a slight 
blue-shift. 
The presence of a red-shifted B component in the subtracted profile in night~3 
suggest that the flare event might have begun this night, 
taking $\sim$~24 hours to reach maximum (night~4). 
This red asymmetry is often
interpreted as the result of chromospheric downward condensations (CDC) 
(Canfield et al.  1990 and references therein).
No significant B component is observed in quiescent state.
The $EW$(H$_\alpha$) returns on night~10 to the same value as observed in the 
quiescent state in night~2 (see Table~\ref{tab:actind}). 
Analyzing this $EW$ data, 
we estimate a duration of approximately six days for this flare event.

In Table~\ref{tab:flares} we list the intensity (I), FWHM and $EW$ 
of the B and N components 
and the shift between them ($\lambda_{\rm N}$~-~$\lambda_{\rm B}$), 
as well as the contribution of B and N components to the total 
$EW$ in the H$\alpha$ line. 
The total values of I, $EW$ and $\log{{\rm F}}$ are also given.

\begin{table*}
\caption[]{Measured $EW$(\AA) of He~{\sc i} D$_{\rm 3}$, 
Na~{\sc i} D$_{\rm 1}$ and
D$_{\rm 2}$ and the Mg~{\sc i} b triplet 
in the subtracted spectra during the flares.
\label{tab:hel_flare}}
\begin{flushleft}
\scriptsize
\begin{tabular}{llccccccccccccc}
\noalign{\smallskip}
\hline
\hline
\noalign{\smallskip}
Date  & UT & He~{\sc i} D$_{\rm 3}$  &  Na~{\sc i} D$_{\rm 1}$  &  Na~{\sc i} D$_{\rm 2}$  & Mg~{\sc i} b$_{\rm 1}$  & Mg~{\sc i} b$_{\rm 2}$ & Mg~{\sc i} b$_{\rm 3}$  &  Fe~{\sc ii} $\lambda$5169 \AA \\
\scriptsize
\\
\noalign{\smallskip}
\hline
\noalign{\smallskip}
\multicolumn{4}{l}{\bf HET-HRS 2001/12} \\
\noalign{\smallskip}
\hline
\noalign{\smallskip}
22 Dec (pre-flare) & 03:44 & - - & 0.087$\pm$0.010 & 0.123$\pm$0.012 & - - & - - & - - & - - \\
23 Dec (flare max) & 04:00 & 0.208$\pm$0.006 & 0.255$\pm$0.006 & 0.273$\pm$0.008 & 0.042$\pm$0.001 & 0.045$\pm$0.001 & 0.062$\pm$0.001 & 0.082$\pm$0.001 \\
26 Dec (gradual decay) & 03:09 & - - & 0.074$\pm$0.010 & 0.102$\pm$0.014 & - - & - - & - - & - - \\
\noalign{\smallskip}
\hline
\noalign{\smallskip}
\multicolumn{4}{l}{\bf NOT-SOFIN 2002/08} \\
\noalign{\smallskip}
\hline
\noalign{\smallskip}
22 Aug (pre-flare) & 04:16 & - - & 0.099$\pm$0.002 & 0.147$\pm$0.002 & - - & - - & - - & - - \\
23 Aug (flare max) & 04:24 & 0.023$\pm$0.001 & 0.112$\pm$0.002 & 0.167$\pm$0.002 & - - & - - & - - & - - \\
24 Aug (quiescent) & 04:39 & - - & 0.093$\pm$0.001 & 0.147$\pm$0.001 & - - & - - & - - & - - \\
%
\noalign{\smallskip}
\hline
\end{tabular}

\end{flushleft}
\end{table*}

\begin{table*}
\caption[]{H$\alpha$ line parameters  
(total and narrow and broad components) 
measured in the subtracted spectra
during the flares
\label{tab:flares}}
\begin{flushleft}
\scriptsize
\begin{tabular}{llcccccccccccccc}
\noalign{\smallskip}
\hline
\hline
\noalign{\smallskip}
 & & \multicolumn{3}{c}{H$\alpha$ Total emission (T)} & &  
\multicolumn{4}{c}{H$\alpha$ Broad component (B)} & & 
\multicolumn{4}{c}{H$\alpha$ Narrow component (N)}\\
\cline{3-5} \cline{7-10} \cline{12-15}\\
Date  & UT & I & EW & $\log{\rm F}$ & &
 I & FWHM & EW & B/T & & 
 I & FWHM & EW & N/T & 
$\Delta \lambda$ \\
 & & & (\AA) & & & & (\AA) & (\AA) & \% & & &  (\AA) & (\AA) & \% & 
$\lambda_{\rm N}$ - $\lambda_{\rm B}$ 
\scriptsize
\\
\noalign{\smallskip}
\hline
\noalign{\smallskip}
\multicolumn{4}{l}{\bf HET-HRS 2001/12} \\
\noalign{\smallskip}
\hline
\noalign{\smallskip}
21 Dec (quiescent)     & 03:37 & 0.75 & 1.16 & 6.478 & &
 -   &  -   &  -   & -  & &  -   &  -   &  -   & -  &  -   \\
22 Dec (pre-flare)     & 03:44 & 0.82 & 1.31 & 6.531 & & 
0.13 & 2.94 & 0.41 & 31 & & 0.72 & 1.18 & 0.90 & 69 & -0.14 \\ 
23 Dec (flare max) & 04:00 & 1.06 & 2.75 & 6.853 & &
0.29 & 5.29 & 1.66 & 60 & & 0.79 & 1.30 & 1.09 & 40 & \ 0.10 \\
26 Dec (gradual decay) & 03:09 & 0.81 & 1.42 & 6.566 & &
0.14 & 2.52 & 0.37 & 26 & & 0.69 & 1.16 & 1.05 & 74 & \ 0.04 \\
\noalign{\smallskip}
\hline
\noalign{\smallskip}
\multicolumn{4}{l}{\bf NOT-SOFIN 2002/08} \\
\noalign{\smallskip}
\hline
\noalign{\smallskip}
22 Aug (pre-flare)     & 04:16 & 0.83 & 1.31 & 6.531 & &
0.06 & 4.98 & 0.31 & 24 & & 0.80 & 1.23 & 1.00 & 76 & \ 0.76\\
23 Aug (flare max) & 04:24 & 0.96 & 1.71 & 6.647 & &
0.09 & 5.16 & 0.48 & 28 & & 0.90 & 1.28 & 1.23 & 72 &  -0.40\\
24 Aug (quiescent)     & 04:39 & 0.81 & 1.29 & 6.524 & &
 -   &  -   &  -   & -  & &  -   &  -   &  -   & -  &    -  \\
%
\noalign{\smallskip}
\hline
\end{tabular}

\end{flushleft}
\end{table*}

\subsection{Description of the flare on 2002 August 23}

The second flare occurred during the NOT-SOFIN 2002/08 observing run 
and appears to be a less energetic or more rapid 
event. The excess H$\alpha$ emission $EW$ in the subtracted
spectra increases in a factor of $\sim$~1.46 from the quiescent state to 
the flare maximum (significantly different from the 2.1 factor obtained in 
HET-HRS 2001/12). 
 
In Table~\ref{tab:hel_flare} we list the measured $EW$ 
of He~{\sc i} D$_{\rm 3}$ and
Na~{\sc i} doublet in the subtracted spectra. 
The Mg~{\sc i} b triplet is not observed  
at this epoch due to a wavelength gap in the configuration of 
the spectrograph.
Emission in the Na~{\sc i} doublet is observed in the subtracted spectra 
in both flare spectra and is stronger at flare maximum. 
The emission in He~{\sc i} D$_{\rm 3}$ is present only 
in the flare maximum spectrum (see Fig.~\ref{fig:flare_na_not}). 
The much lower $EW$(He~{\sc i} D$_{\rm 3}$) and the increment in 
$EW$(Na~{\sc i} D1 \& D2) measured at flare maximum both indicate 
a less energetic event than that observed on 2001 December 23.

As in the case of HET-HRS 2001/12 run, a broad emission component (B) in both
the subtracted and observed H$\alpha$ profile is present at flare maximum 
(see Fig.~\ref{fig:flare_ha_not}). 
The two-Gaussian fit used yields a FWHM $\sim$ 236 km s$^{\rm -1}$ for the 
B component and FWHM $\sim$ 58 km s$^{\rm -1}$ for the narrow one (N).
In the pre-flare phase we obtain a FWHM for the N component of 
56 km s$^{\rm -1}$, very similar to that obtained for N component in 
HET-HRS 2001/12 (53 km s$^{\rm -1}$). 
However, the FWHM of 228 km s$^{\rm -1}$ 
in the B component is very different of 134 km s$^{\rm -1}$ 
from HET-HRS 2001/12. 
No significant B component is detected on the night~3; 24 Aug.  
This suggests that flare maximum likely occurred between night~1 and night~2 
and that the maximum observed on night~2 is part of the gradual decay phase.
Thus, the duration of this event is between 2 and 3 days, 
shorter than in 2001/12.
Contrary to what is observed in the HET-HRS 2001/12 event, 
the B component is blue-shifted in the pre-flare observation 
and red-shifted at the maximum (night~2). 
Different dynamic processes could be taking place during this flare event. 
Another possible explanation is that these asymmetries could be due
the position of the active region over the disk. 

In Table~\ref{tab:flares} we list the H$\alpha$ line parameters 
I, FWHM, $EW$, and relative contribution of the B and N components,
as well as the shift between them ($\lambda_{\rm N}$~-~$\lambda_{\rm B}$) 
during the flare.
The total values of I, $EW$, and $\log{\rm F}$ of the total 
excess H$\alpha$ emission are also presented.


\section{Photospheric variations}

Since PW And is a young K2-dwarf with a rotational velocity of
22.6 km s$^{\rm -1}$, it is expected to show large dark photospheric
spots which produce rotational modulation of the light coming from its surface.
A photometric period was obtained by Hooten \& Hall (1990)
by fitting a sine curve to the data using a standard less squares technique. 
Values of 1.64 $\pm$ 0.01, 29 $\pm$ 1 and 1.745 $\pm$ 0.005 days were 
calculated for three different epochs.  
They propose 1.745 days as the more reliable period is it had the least 
internal scatter.

To study spectroscopic variations produced by spots we have used the 
fact that cool spots moving across the disk of a star will produce a 
weaker contribution to the total integrated line profile.  
Thus the spectra -integrated over the disk will show
a bump traveling across the profile as the star rotates.  
Two methods are commonly used to interpret this information. 
Doppler imaging was developed with the aim of 
obtaining a two dimensional picture of spots (Vogt \& Penrod 1983,
Vogt et al. 1987, Collier-Cameron \& Unruh 1994, Rice et al. 1989, 
Strassmeier \& Rice 1989). Here high signal-to-noise ratio (S/N) and resolution 
($\lambda$/$\Delta \lambda \sim 100000$) are needed in order to 
obtain both accurate position and dimensional information on the spot(s). 
Another method, first used in the case of the Sun, can be
adapted to stars using high $\lambda$/$\Delta \lambda$ and S/N spectra
(Toner \& Gray 1988). The technique consists of calculating
a bisector for line profile, i.e.
the middle points of the line profile taking points of equal intensity in 
both sides of the line. Variations of the bisector are related
to the existence of surface features moving across the disk, as seen 
in $\xi$~Boo~A by Toner \& Gray (1988) 
and in $\lambda$~And by Donati et al. (1995). 
Both techniques have serious limitations (Dempsey 1991). 
Doppler imaging requires the star to be rotating rapidly 
$v\sin i > 36$ km s$^{\rm -1}$ to give sufficient spatial resolution. 
Values of $i$ over 10-30 degrees are needed to guarantee high
rotational broadening. However, the distortions induced by spots are 
of the order of 1-10 per cent of the continuum, and S/N $\ge$ 300 
is required to detect them.
Moreover, bisector analysis is only appropriate for very sharp
lines, $v\sin i < 10$ km~s$^{\rm -1}$, restricting the candidates to
the very slow rotators or stars with very low value of inclination.
S/N $\ge$ 150 is also needed to apply this technique. 
A common problem with both techniques is the use of only one unblended 
line for the study, although several lines can be combined in order to 
obtain more accurate results.

\begin{figure}
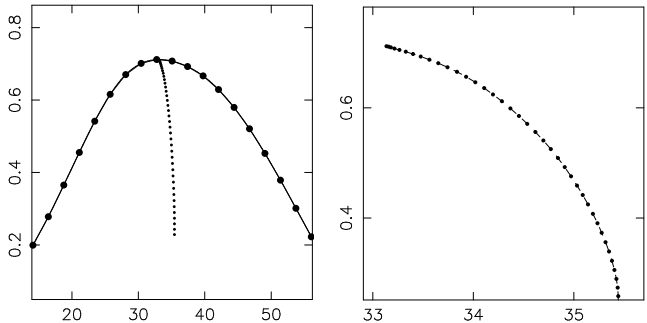

\centerline{{\psfig{figure=MS3636f9.ps,width=4.3cm,bbllx=650pt,bblly=280pt,bburx=850pt,bbury=490pt,clip=}}
{\psfig{figure=MS3636f9.ps,width=4.3cm,bbllx=650pt,bblly=72pt,bburx=850pt,bbury=285pt,clip=}}}
\caption[ ]{{\bf Left}: Example of a cross-correlation function (CCF). 
Abscissa is given in units of pixels.
The peak of the CCF has been fitted by a cubic-spline. 
{\bf Right}: A zoom of the bisector is plotted in the right panel.
\label{fig:ccfbisector}}
\end{figure}

\begin{table}
\caption[]{Heliocentric radial velocities and CCF bisector parameters
\label{tab:vhel}}
\begin{center}
\scriptsize
\begin{tabular}{lcccccccccccccc}
\noalign{\smallskip}
\hline
\hline
\noalign{\smallskip}
\tiny Date  & \tiny UT & $V_{\rm hel}$ & $\Delta \lambda_{\rm bis}$ & $\Delta v_{\rm b}$ & 
Phase$^{a}$\\
                 & (hh:mm)  & (km s$^{-1}$) & (\AA) & (km s$^{-1}$) \\
\scriptsize
\\
\noalign{\smallskip}
\hline
\noalign{\smallskip}
\multicolumn{3}{l}{\bf 2.2m-FOCES 1999} \\
\noalign{\smallskip}
\hline
\noalign{\smallskip}
25 Jul & 04:09 & -11.76 $\pm$ 0.59 & \ 0.043 & \ 1.90 & 0.00\\
26 "   & 00:57 & -11.49 $\pm$ 0.23 &  -0.007 &  -0.35 & 0.49\\
27 "   & 02:25 & -12.50 $\pm$ 0.37 & \ 0.070 & \ 3.14 & 0.10\\
28 "   & 00:57 & -10.51 $\pm$ 0.42 &  -0.044 &  -1.97 & 0.64\\
29 "   & 03:01 & -11.95 $\pm$ 0.26 & \ 0.040 & \ 1.78 & 0.26\\
30 "   & 01:57 & -11.11 $\pm$ 0.47 & \ 0.037 & \ 1.63 & 0.81\\
\noalign{\smallskip}
\hline
\noalign{\smallskip}
\multicolumn{3}{l}{\bf HET-HRS 2001-2002} \\
\noalign{\smallskip}
\hline
\noalign{\smallskip}
20 Dec & 04:12 & -13.08 $\pm$ 0.41 & \ 0.038 & \ 1.72 & 0.73\\
21  "  & 03:37 & -13.77 $\pm$ 0.30 & \ 0.014 & \ 0.65 & 0.29\\
22  "  & 03:44 & -12.45 $\pm$ 0.35 &  -0.022 &  -0.99 & 0.86\\
23  "  & 04:00 & -11.51 $\pm$ 0.34 &  -0.043 &  -1.96 & 0.44\\
26  "  & 03:09 & -12.40 $\pm$ 0.40 & \ 0.046 & \ 2.07 & 0.14\\
27  "  & 03:07 & -14.83 $\pm$ 0.35 & \ 0.043 & \ 1.95 & 0.71\\
28  "  & 03:31 & -12.34 $\pm$ 0.30 &  -0.014 &  -0.64 & 0.29\\
29  "  & 02:52 & -15.04 $\pm$ 0.46 &  -0.001 &  -0.07 & 0.85\\
30  "  & 02:59 & -11.90 $\pm$ 0.33 &  -0.038 &  -1.71 & 0.43\\
\noalign{\smallskip}
\hline
\noalign{\smallskip}
\multicolumn{3}{l}{\bf NOT-SOFIN 2002} \\
\noalign{\smallskip}
\hline
\noalign{\smallskip}
22 Aug & 04:16 &  -11.62 $\pm$ 0.28 & \ 0.047 & \ 2.09 & 0.13\\
23 "   & 04:24 &  -11.65 $\pm$ 0.34 & \ 0.042 & \ 1.86 & 0.71\\
24 "   & 04:39 &  -11.73 $\pm$ 0.25 & \ 0.055 & \ 2.44 & 0.29\\
25 "   & 01:55 &  -11.96 $\pm$ 0.33 & \ 0.108 & \ 4.84 & 0.79\\
26 "   & 05:15 & \ -8.83 $\pm$ 0.26 &  -0.054 &  -2.43 & 0.45\\
26 "   & 05:37 & \ -8.81 $\pm$ 0.32 &  -0.034 &  -2.15 & 0.45\\
27 "   & 05:06 & \ -9.95 $\pm$ 0.29 & \ 0.014 & \ 0.60 & 0.02\\
28 "   & 04:19 & \ -9.03 $\pm$ 0.40 &  -0.031 &  -1.42 & 0.57\\
29 "   & 02:37 &  -10.61 $\pm$ 0.34 & \ 0.015 & \ 0.67 & 0.10\\
%
\hline
\noalign{\smallskip}
\end{tabular}

\end{center}
\hspace{0.3cm}
{\scriptsize
$^{\rm a}$ $P_{\rm phot} = 1.745$ days (from Hooten \& Hall 1990).}
\end{table}

A powerful method was developed and applied to several stars in the 
range of $15 < v \sin i < 40$ km s$^{\rm -1}$
by Dempsey et al. (1992) using a correlative analysis. A 
non active star (template) is cross-correlated with the active star producing
a cross-correlation function (CCF). Variations in the peak of the CCF are 
related to changes in the line profiles caused by spots. To quantify the 
temporal variations in the CCF the bisector of the peak of the CCF can 
be calculated. This CCF bisector is not the same as the bisector 
described by Toner \& Gray (1988), since the bisector of a line
measures velocity fields while the CCF bisector quantifies the asymmetry of 
the CCF.
The best results are 
obtained when both template and active stars have a similar spectral type.
The advantage of this technique is the possibility of using many absorption 
lines for the calculation of the CCF (Dempsey et al. 1992). As the information
of the lines is redundant, there is less restriction in the 
S/N ($\sim$150). Resolution of $\lambda$/$\Delta \lambda \sim$ 40000 is
sufficient to obtain accurate results. We employ this latter technique to
study the effects of star-spots on the profiles of photospheric
lines in PW And.

Three observing runs with the most extensive rotational phase 
coverage are used to study the effects of photospheric spots.  
These are the 2.2m-FOCES 1999/07, HET-HRS 2001/12 
and NOT-SOFIN 2002/08 observations (see Section 2 and Table~\ref{tab:obslog}). 
The CCFs were determined using the IRAF routine {\sc fxcor} 
in the same procedure as used in the determination of the radial 
velocities $V_{\rm hel}$ (see Section 3.2).  
Here we limit calculations to  
the wavelength regions ranging from 6300 to 6465~\AA \ 
and 6670 to 6760~\AA \ (see Fig.~\ref{fig:f99_lineas} 
and ~\ref{fig:not02b_lineas}), 
which includes lines commonly used in Doppler imaging 
like Fe~{\sc i} lines in 6411.644 and 6430.841~\AA \ 
and Ca~{\sc i} lines in 6439.073, 6462.566 and 6717.680~\AA. 
The asymmetries of the CCF are very subtle in PW And necessitating 
calculation of its bisector.
The determination of the CCF bisector has been carried out in the usual 
manner, taking points of equal intensity for the calculation of the
middle points.
For each point on the left of the CCFs, 
a matching point is found on the right using a cubic-spline interpolation 
(see Fig.~\ref{fig:ccfbisector}).
In order to quantify the changes in the CCF bisector, the difference
between the mean of ten interpolated points in the top and in the base 
of the CCF bisector ($\Delta \lambda_{\rm bis}$) has been measured.
We have choose a fixed value of 0.2 as the base of the CCF
(see Fig.~\ref{fig:ccfbisector}) to avoid the noise at the zero level of the CCF.
In the subsequent discussion the values of $\Delta \lambda_{\rm bis}$ have been 
converted into velocities ($\Delta v_{\rm b}$) for better understanding the science.

We have analyzed the CCF of each observing run separately,
in order to avoid the possible slightly differences of the instrumental
profile of the different spectrographs, 
and because the behaviour of the photospheric activity could be different
at these three epochs separated by nearly three years.

\subsection{CCF bisectors in the observing run 2.2m-FOCES 1999/07}

\begin{figure}
{\psfig{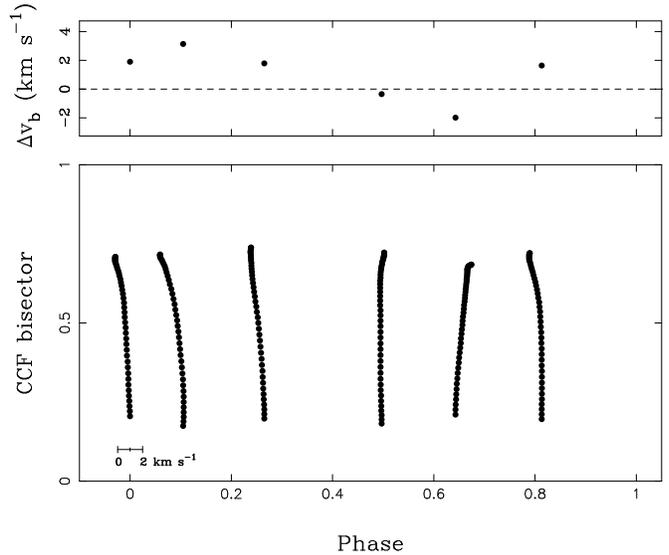}}
\caption[ ]{CCF bisectors (lower panel) and $\Delta v_{\rm b}$ 
(upper panel) 
for the observations of 2.2m-FOCES 1999/07 run arranged in phase using
the photometric period given by Hooten \& Hall (1990) and the 
first observation as phase 0.0. 
\label{fig:f99_bisec}}
\end{figure}

The six observations over six nights cover
almost three rotational cycles.
%
GJ~706 (K2 V) has been used as template
to calculate the CCF for PW And. 
In Fig.~\ref{fig:f99_bisec} we have plotted in phase
the CCF bisectors and the values of 
%
$\Delta v_{\rm b}$
using a period of $P = 1.745$ days from Hooten \& Hall (1990) with
the time of the first observation ($T_{\rm o}$) taken as phase 0.0.
%

\begin{figure}
{\psfig{figure=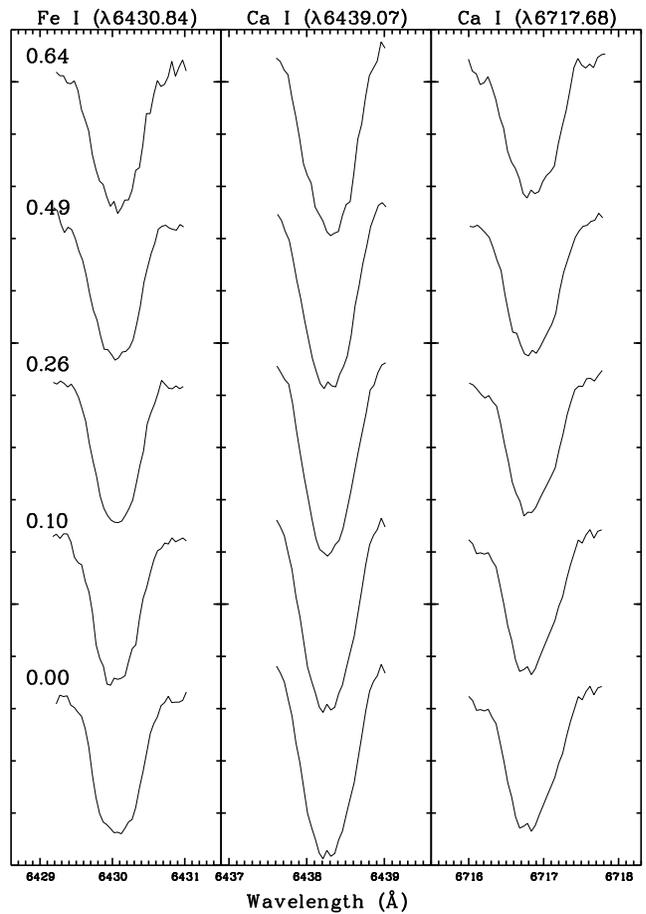,width=8.7cm,bbllx=45pt,bblly=28pt,bburx=425pt,bbury=565pt,clip=}}
\caption[ ]{Line profiles ($\lambda \lambda$6430.84, 6439.07 
and 6717.68 \AA)
of PW And obtained during 2.2m-FOCES 1999/07.
Phase is given at the left panel, for each observation. 
The same pattern is seen in the three lines for each phase.
\label{fig:f99_lineas}}
\end{figure}

The maximum $\Delta v_{\rm b}$ measured is 3.14 km s$^{-1}$
(see Table~\ref{tab:vhel}) at phase 0.1, changing to
-1.97 km s$^{-1}$ at phase 0.64.
The sequence is compatible with the existence of two spots 
at nearly opposite longitudes, one of them near the pole since a perturbation 
in the core of the line is present in every phase. 
In Fig.~\ref{fig:f99_lineas} we have plotted the 
profile of Fe~{\sc i}~6430.84~\AA \ and Ca~{\sc i}~6439.07 and 6717.68~\AA \ 
photospheric lines. A disturbance in the core of the Ca~{\sc I} lines, 
slightly shifted with phase, is clearly present for all phases. 

\begin{figure}
{\psfig{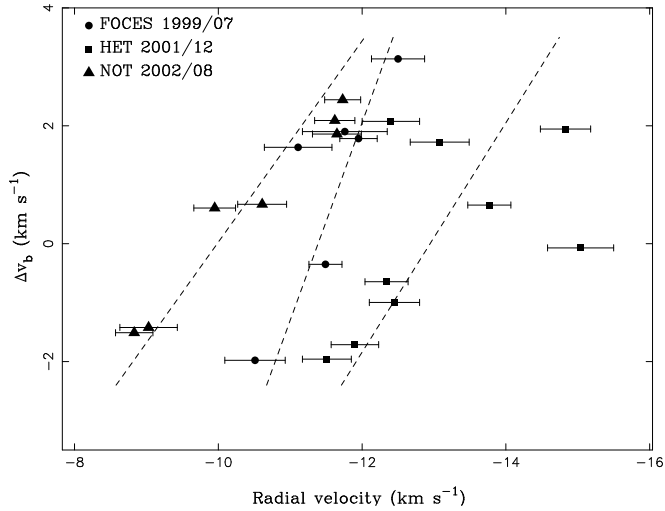}}
\caption[ ]{$\Delta v_{\rm b}$ in the bisector vs. $V_{\rm hel}$.
Changes in $V_{\rm hel}$ are related to the existence of spots on the star.
Differences in $V_{\rm hel}$ between the different observing runs could be 
due to the different radial velocity
standard stars used. 
Changes in the slope of the lines
could be due to changes in the pattern of spots.
\label{fig:vr_correl}}
\end{figure}

The variations observed in the radial velocity ($V_{\rm hel}$), 
determined as is explained in Section 3.2, 
are related to the existence of spots on the star.
A clear correlation between 
$\Delta v_{\rm b}$
and $V_{\rm hel}$
is evident (see Fig.~\ref{fig:vr_correl}).

\subsection{CCF bisectors in the observing run HET-HRS 2001/12}

Nine observations of PW And were obtained during this observing run 
covering 6.3 rotational periods 
with a gap of two days between nights~4 and 7.
In order to obtain accurate results for CCF bisectors, the non-active star
GJ~706 from 2.2m-FOCES 1999/07 was used as template. The spectrum of 
GJ~706 was convolved with a Gaussian in order to degrade the spectral 
resolution to the lower $\lambda$/$\Delta \lambda$ in 
HET-HRS 2001/12.

\begin{figure}
{\psfig{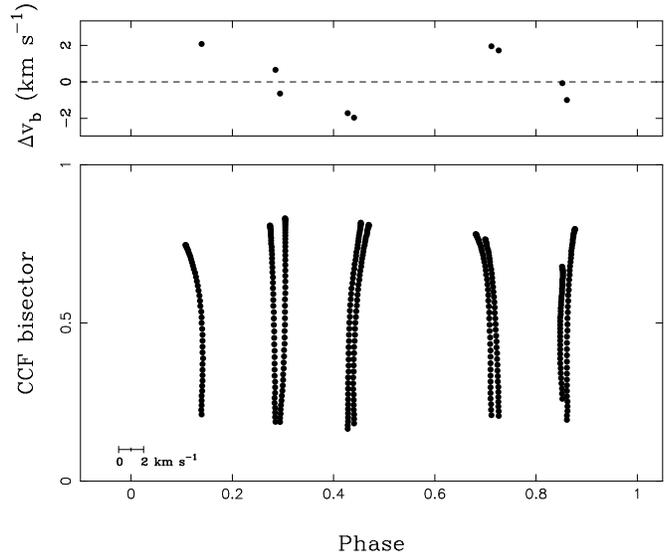}}
\caption[ ]{CCF bisectors and $\Delta v_{\rm b}$ 
for the observations of HET-HRS 2001/12 arranged in phase using
the photometric period given by Hooten \& Hall (1990) and $T_{\rm o}$ 
of 2.2m-FOCES 1999/07 as phase 0.0.
\label{fig:het01_bisec}}
\end{figure}

CCF bisectors have been plotted in Fig.~\ref{fig:het01_bisec}, using 
the same $T_{\rm o}$ and $P_{\rm phot}$ as in 
2.2m-FOCES 1999/07 with the aim of comparing
both observing runs. In spite of the more than two years time difference,
the change in the CCF bisector follows a sequence very similar to that for 
2.2m-FOCES 1999/07 when arranged in phase. 
This could be due to the existence of large patterns of spots persisting
with time, perhaps due to active longitudes.

As in the previous observing run,
a correlation between 
$\Delta v_{\rm b}$
and $V_{\rm hel}$ is found (see Fig.\ref{fig:vr_correl}).
In this case the slope is different and, due to the lower spectral
resolution of this run, a larger scatter in $V_{\rm hel}$
is observed.

\subsection{CCF bisectors in the observing run NOT-SOFIN 2002/08}

\begin{figure}
{\psfig{figure=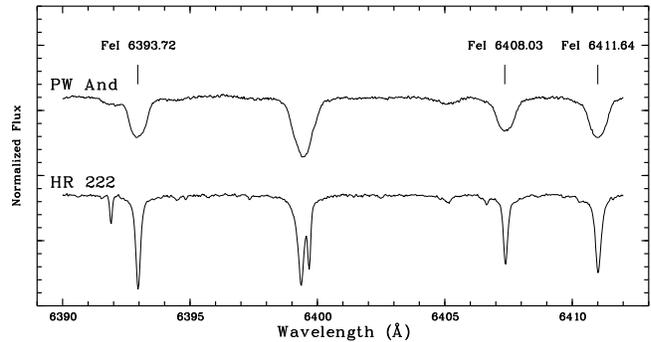,width=8.7cm,bbllx=28pt,bblly=28pt,bburx=550pt,bbury=320pt,clip=}}
\caption[ ]{Observed spectrum of PW And and HR 222 in the
region between 6360 to 6415~\AA \ used in the determination of 
the CCF in NOT-SOFIN 2002/08. 
The blended $\lambda \lambda$ 6400.33 and 6400.01
have been rejected in this work.
\label{fig:ref_lineas}}
\end{figure}

For this data set we used the template K2-dwarf star 
HR 222 (observed during every night) as both the
radial velocity and spectral type standard in calculating the CCF.
Due to a large gap in the spectral region between 6420 and 6540~\AA \ 
caused by the configuration used here, a range from 6360 to 6415~\AA \ has 
been used for measuring CCF bisectors instead of that employed in the previous 
observing runs. In this region the strong unblended Fe~{\sc i} lines 
$\lambda \lambda$ 6393.72, 6408.03 and 6411.64 \AA \ are present.
The Fe~{\sc i} lines $\lambda \lambda$ 6400.33 and 6400.01 are blended 
and have been rejected for this analysis (see Fig.~\ref{fig:ref_lineas}).

As in the preceding cases, the photometric period of 1.745 days has been 
used to calculate the phase of the CCF bisectors. 
The results have been plotted using $T_{\rm o}$ of 
2.2m-FOCES 1999/07 as phase 0.0 (Fig.~\ref{fig:not02b_bisec}).

A clear difference in the phase between this and the previous observing runs 
is present. Nevertheless, the sequence is similar to that in HET-HRS 2001/12
and 2.2m-FOCES 1999/07.
{The maximum $\Delta v_{\rm b}$ measured is 4.84 km s$^{-1}$
(see Table~\ref{tab:vhel})
at phase 0.79. This is larger than the obtained on July 1999, and
less than 2.50 km s$^{-1}$ is measured for the rest of the phases.
This suggests the 
presence of a large spot on the approaching limb.

The spot disturbance in the line profile is not the same 
as in 2.2m-FOCES~1999/07.
The presence of a perturbation in the core of absorption lines is not clear 
in Fig.~\ref{fig:not02b_lineas}, although a feature disturbing the profile is 
shown. There is no apparent evidence for a polar spot.

The correlation between 
$\Delta v_{\rm b}$
and $V_{\rm hel}$
is plotted in Fig.\ref{fig:vr_correl}.
In this case the slope of the line is similar to that of
HET-HRS 2001/12 observing run. 
These changes in the slope from one observing run to another could 
indicate changes in the pattern of the spots at these epochs.

\begin{figure}
{\psfig{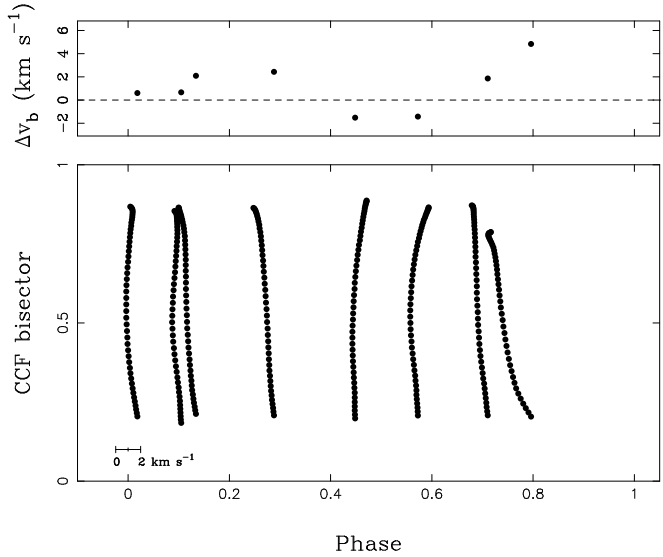}}
\caption[ ]{CCF bisectors and $\Delta v_{\rm b}$ 
for the observations of NOT-SOFIN 2002/08 arranged in phase using
the photometric period given by Hooten \& Hall (1990) and $T_{\rm o}$
of 2.2m-FOCES 1999/07 as phase 0.0.
\label{fig:not02b_bisec}}
\end{figure}

\begin{figure}
{\psfig{figure=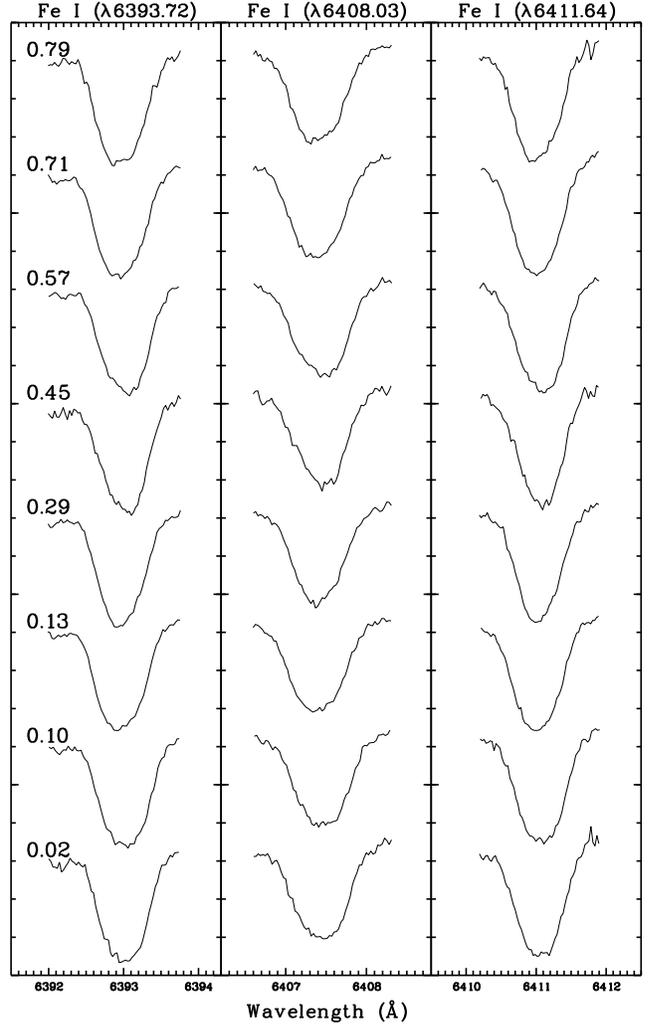,width=8.7cm,bbllx=45pt,bblly=28pt,bburx=425pt,bbury=635pt,clip=}}
\caption[ ]{Line profiles ($\lambda \lambda$ 
6393.72, 6408.03 and 6411.64 \AA)
of PW And obtained during NOT-SOFIN 2002/08.
Phase is given at the left panel, for each observation.
\label{fig:not02b_lineas}}
\end{figure}

\section{Relation between photospheric and chromospheric activity}

When studying the chromospheric activity indicators in Section 4,  
variations in the $EW$ of several lines, such as H$\alpha$, H$\beta$ and 
Ca~{\sc ii} H~\&~K were found (see Tables~\ref{tab:actind} \& 
\ref{tab:actflux}). 
When arranged in phase, the $EW$(H$\alpha$) and $EW$(Ca~{\sc ii} H \& K) 
appear to follow a clear sequence for each observing run.
Comparison with results from the CCF bisectors shows a correlation 
between photospheric and chromospheric activity. 
In Figs.~\ref{fig:f99_vract}, \ref{fig:het01_vract} \& \ref{fig:not02b_vract}
we have over-plotted $\Delta \lambda_{\rm bis}$ and $V_{\rm hel}$
with the $EW$ of chromospheric lines for each observing run.

\subsection{2.2m-FOCES 1999/07}

For this run we study the chromospheric lines 
H$\alpha$ and Ca~{\sc ii} H \& K 
in relation to the photospheric line behavior (see Fig.~\ref{fig:f99_vract}). 
The $EW$(Ca~{\sc ii} H) and $EW$(Ca~{\sc ii} K) have been summed together due to the low S/N of the continuum in this region of our spectra
(see Table~\ref{tab:obslog}). 
Note that the $EW$ of Ca~{\sc ii} lines could not be measured on 
night~4, which corresponds to phase 0.64. 

A clear relation between $EW$(Ca~{\sc ii}~H~+~K) and
the variations in 
$\Delta v_{\rm b}$
and $V_{\rm hel}$ is observed 
(see Fig.~\ref{fig:f99_vract}).
For the H$\alpha$ line, the relation is not as clear, 
but a similar trend is followed from phase 0.5 to 1.0.  
This supports the existence of active regions 
connected to the photospheric features. 
The presence of plage-like regions can be inferred from the variation on 
Ca~{\sc ii} H \& K lines emission and this conjecture is further supported by the values of $E_{8542}$/$E_{8498}$ 
calculated in Section 5 and listed in Table~\ref{tab:actflux}.

\begin{figure}
{\psfig{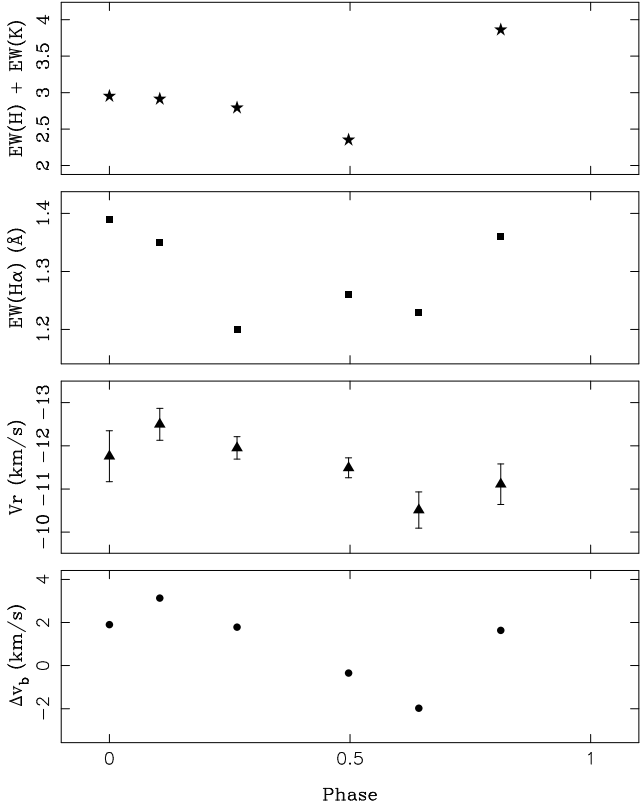}}
\caption[ ]{Comparison between photospheric ($\Delta v_{\rm b}$, $V_{\rm hel}$)
and chromospheric ($EW$(H$\alpha$), $EW$(Ca~{\sc ii} H \& K)) variations in 
2.2m-FOCES 1999/07.
\label{fig:f99_vract}}
\end{figure}

\subsection{HET-HRS 2001/12}

The chromospheric lines H$\alpha$ and Ca~{\sc ii} $\lambda$8542 \AA \ are used 
for the comparison during this run since Ca~{\sc ii} H \& K lines 
were beyond wavelength range observed in this epoch (see Section 2).

In Fig.~\ref{fig:het01_vract} we have plotted the $EW$ of these lines 
as well as CCF bisectors and $V_{\rm hel}$. 
The Ca~{\sc ii} $\lambda$8542 \AA \ line does not 
appear to be a good indicator of changes in the activity of the star 
at this epoch. 
Only the flare on 2001 December 23, corresponding to phase 0.44, 
is clearly visible using this line. 
Taking into account that $V_{\rm hel}$ is a minimum at this phase, the
flare occurred when the photospheric feature was near the limb. 

Variations in H$\alpha$ follow a pattern similar to that found 
in CCF bisectors and $V_{\rm hel}$
only up to phase 0.5. 
The less more patchy phase sampling in this data set makes any 
correlation less clear.

\begin{figure}
{\psfig{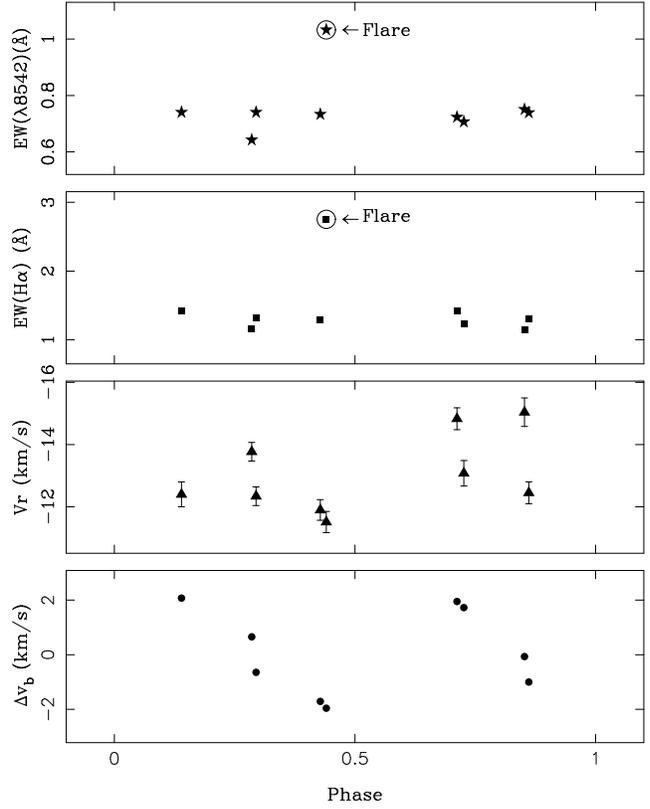}}
\caption[ ]{Comparison between photospheric and chromospheric variations in
HET-HRS 2001/12.
\label{fig:het01_vract}}
\end{figure}

\subsection{NOT-SOFIN 2002/08}

For this observing run we have again the H$\alpha$ and Ca~{\sc ii} K lines. 
The flare on 2003 August 23 corresponds to phase 0.71 and is clearly visible in 
both chromospheric lines (see Fig.~\ref{fig:not02b_vract}). 
Contrary to the event in
HET-HRS 2001/12, the flare does not appear to have occurred over the limb, 
but seems more associated with an active region
on the disk. 
An overall correlation between both H$\alpha$ and Ca~{\sc ii} K lines 
with CCF bisectors and $V_{\rm hel}$ is apparent. 
This suggests that the chromospheric active 
regions are connected to the photospheric features in the same way 
as 2.2m-FOCES 1999/07, 
although the correlation of H$\alpha$ is perhaps more clear at this epoch. 

In order to look for any features in the H$\alpha$ emission line that 
are related to the variations seen in the photospheric lines, 
we have applied the bisectors method to the emission profile in
the subtracted spectrum (see Fig.~\ref{fig:not02b_Ha}). 
Again, for each point on the left of the H$\alpha$ subtracted profile, 
a matching point on the right is found with a cubic-spline interpolation 
to obtain a bisector of the subtracted emission line. 
Any features moving across the profile must disturb this bisector in the 
same way as in the case of a CCF bisector. 
The resulting profile bisectors and 
$\Delta v_{\rm b}$
are plotted in Fig.~\ref{fig:Ha_bisec}, using the same 
$P_{\rm phot}$ and $T_{\rm o}$ as in Fig.~\ref{fig:not02b_bisec}. 
The scale of the bisectors has been changed to make it easier to see 
the curvature. Only the top of the
profile is perturbed by the features, while the bottom is less sensitive, 
probably due to the difference in flux and S/N between the core 
and the wings of the emission.
In order to avoid changes in the bottom of the profile bisector induced by 
changes in the broad component, as is present in the flare state, 
we have discarded $\sim$20 per cent of the bisector nearest the continuum. 
The sequence obtained in Fig.~\ref{fig:Ha_bisec} is
similar to that of Fig.~\ref{fig:not02b_bisec} suggesting the presence 
of cool chromospheric material associated with the photospheric features.  
This is consistent with the results 
of the $EW$ of chromospheric emission lines (see Fig.~\ref{fig:not02b_vract}).

\begin{figure}
{\psfig{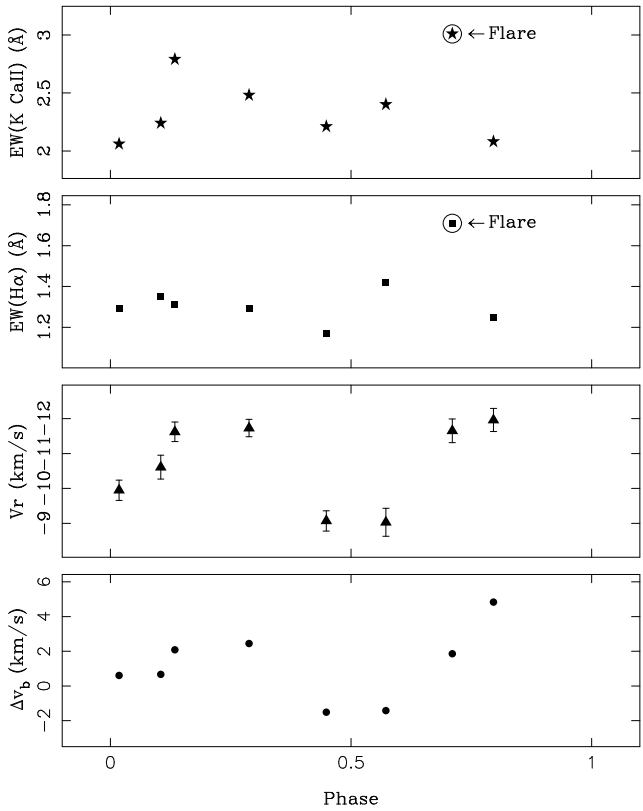}}
\caption[ ]{Comparison between photospheric and chromospheric variations in
NOT-SOFIN 2002/08.
\label{fig:not02b_vract}}
\end{figure}

\begin{figure}
{\psfig{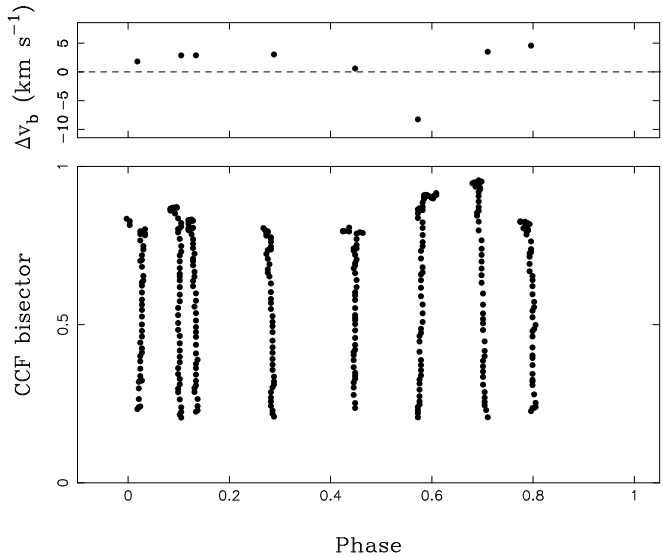}}
\caption[ ]{Bisectors of the subtracted emission line profiles of H$\alpha$ 
for the observations of NOT-SOFIN 2002/08. 
\label{fig:Ha_bisec}}
\end{figure}

\begin{figure}
{\centerline{\psfig{figure=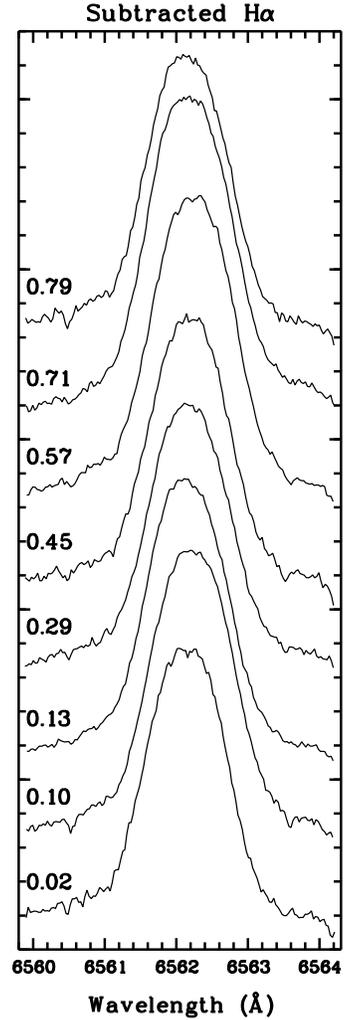,bbllx=45pt,bblly=28pt,bburx=203pt,bbury=425pt,clip=}}}
\caption[ ]{Subtracted H$\alpha$ spectra for the NOT-SOFIN 2003/08 observing run
ordered in phase. Flare occurred for phase 0.71.
\label{fig:not02b_Ha}}
\end{figure}

\section{Conclusions}

In this paper we have analyzed high resolution echelle observations,
taken during the period 1999--2002,
to describe in detail the activity on the young K2-dwarf star PW And. 
We also have presented new determinations of key physical parameters. 
In addition, a study of the lines profile
has been carried out in order to relate the variations found in the 
heliocentric radial velocity ($V_{\rm hel}$) and chromospheric emission 
with the existence of 
photospheric features co-rotating with the star.

Radial velocity standard stars have been used to calculate accurate 
heliocentric radial velocities for each observation by using the
cross-correlation technique. 
Variations in the $V_{\rm hel}$ up to 3~$km~s^{\rm -1}$ 
in the same observing run were found. 
The mean value of the velocities of all the observing runs 
(-11.15 $\pm$ 0.05 $km s^{\rm -1}$)
is used to calculate the Galactic space-velocity components ($U$,$V$,$W$) 
and confirms its membership to the Local Association moving group.
By using the information provided by the width (FWHM) of the peak of the 
cross-correlation function (CCF), we have determined a projected 
rotational velocity, $v \sin i$, of 22.6 $km s^{\rm -1}$. 
The radius for a K2V star ($R = 0.80 \ R_{\odot}$)
is compatible with the minimum radius ($R = 0.78 \pm 0.11 $ R$_{\odot}$) 
obtained from $v \sin i$ and the photometric period.
The age of PW And has been bounded between 30 to 80 Myrs from its position 
in relation to pre-main sequence isochrones in the color 
magnitude diagram.   
The measured $EW$(Li~{\sc I}) of 273~m\AA \ indicates an age similar to that 
of the stars in the Pleiades cluster or younger, which is in agreement with 
the estimate from the color-magnitude diagram. 

Using the spectral subtraction technique, all the optical activity indicators 
from the Ca~{\sc ii} H \& K to Ca~{\sc ii} IRT lines have been analyzed.   
Strong emission in the observed spectra is present 
in H$\alpha$, Ca~{\sc ii} H \& K and IRT, 
the remainder of the Balmer lines are clearly visible in emission in the 
subtracted spectra. 
The ratio $\frac{EW({\rm H\alpha})}{EW({\rm H\beta})}$ that we have found 
in 2.2m-FOCES 1999/07 and NOT-SOFIN 2002/08 suggest 
that the emission of these lines could arise from prominence-like material, 
whereas the ratio of the excess emission $\frac{E_{8542}}{E_{8498}}$ indicates
that the Ca~{\sc ii} IRT emission arise from plage-like regions.
Two flares were detected in different epochs (2001 and 2002) which 
confirms the presence of strong magnetic activity on the star.
At the same time, variations in $EW$(H$\alpha$) and $EW$(Ca~{\sc ii} H \& K) 
are observed in all of the observing runs further confirming magnetic 
surface activity. 

A detailed analysis of the profiles of the photospheric absorption lines 
have been done by calculating the bisectors of the CCF resulting of
cross-correlate the spectra of PW And with the spectrum of 
a non active star of similar spectral type.
These CCF bisectors change with a period similar to the photometric period
of the star, suggesting the presence of features (probably cool dark spots) 
in the photosphere which are co-rotating with the star. 
The variations of the CCF bisectors found
in the three epochs analyzed confirm the prevalence over time of large spots or 
the existence of active longitudes were spots are continuously being generated. 
The chromospheric active regions appear to be associated to the 
photospheric features, as a clear correlation between the variations 
observed in chromospheric activity and in the photospheric lines is found. 
Finally, variations in the bisector of the subtracted H$\alpha$ line 
further support the presence of chromospheric active regions, 
related to the cool photospheric spots.


\begin{acknowledgements}

We would like to thank M.Cruz G\'alvez for help in the reduction
of the observations, Dr. B.H. Foing for allowing us to use the
ESA-MUSICOS spectrograph at Isaac Newton Telescope, 
and Dr.~I.~Ilyn for help with SOFIN spectrograph during the 
observations at the Nordic Optical Telescope (NOT).
This work was supported by the Universidad Complutense de Madrid and
the Spanish Ministerio de Ciencia y Tecnolog\'{\i}a (MCYT), Programa 
Nacional de Astronom\'{\i}a y Astrof\'{\i}sica under grant AYA2001-1448.

\end{acknowledgements}
%
%

\listofobjects


\end{document}